\documentclass[10pt,twocolumn,letterpaper]{article}

\usepackage[pagenumbers]{cvpr} 
\usepackage{multirow}
\usepackage{lipsum}

\usepackage[dvipsnames, table]{xcolor}

\definecolor{myellow1}{rgb}{0.94, 0.8, 0.0}
\colorlet{mygray}{white!15}
\colorlet{myellow}{lightgray!15}

\definecolor{cvprblue}{rgb}{0.21,0.49,0.74}
\usepackage[pagebackref,breaklinks,colorlinks,citecolor=cvprblue]{hyperref}

\def\paperID{6162} 
\def\confName{CVPR}
\def\confYear{2024}

\newcommand\blfootnote[1]{%
\begingroup
\renewcommand\thefootnote{}\footnote{#1}%
\addtocounter{footnote}{-1}%
\endgroup
}

\title{MobileUtr: Revisiting the relationship between light-weight CNN and Transformer for efficient medical image segmentation}

\author{Fenghe Tang\textsuperscript{1\dag} \quad
	Bingkun Nian\textsuperscript{2\dag}\quad
	Jianrui Ding\textsuperscript{3}\quad
        Quan Quan\textsuperscript{4}\quad
        Jie Yang\textsuperscript{2}\quad
        Wei Liu\textsuperscript{2$^{*}$}\quad
	S.Kevin Zhou\textsuperscript{1,4$^{*}$} \\
  $^1$ School of Biomedical Engineering \& Suzhou Institute for Advanced Research, \\University of Science and Technology of China \\
  $^2$ Institute of Image Processing and Pattern Recognition, Shanghai Jiao Tong University \\
  $^3$ School of Computer Science and Technology, Harbin Institute of Technology \\
  $^4$ Key Lab of Intelligent Information Processing of Chinese Academy \\ of Sciences (CAS), Institute of Computing Technology\\
}

\begin{document}
\maketitle
\begin{abstract}
Due to the scarcity and specific imaging characteristics in medical images, light-weighting Vision Transformers (ViTs) for efficient medical image segmentation is a significant challenge, and current studies have not yet paid attention to this issue. This work revisits the relationship between CNNs and Transformers in lightweight universal networks for medical image segmentation, aiming to integrate the advantages of both worlds at the infrastructure design level. In order to leverage the inductive bias inherent in CNNs, we abstract a Transformer-like lightweight CNNs block (ConvUtr) as the patch embeddings of ViTs, feeding Transformer with denoised, non-redundant and highly condensed semantic information. Moreover, an adaptive Local-Global-Local (LGL) block is introduced to facilitate efficient local-to-global information flow exchange, maximizing Transformer's global context information extraction capabilities. Finally, we build an efficient medical image segmentation model (MobileUtr) based on CNN and Transformer. Extensive experiments on five public medical image datasets with three different modalities demonstrate the superiority of MobileUtr over the state-of-the-art methods, while boasting lighter weights and lower computational cost. Code is available at \url{https://github.com/FengheTan9/MobileUtr}.\blfootnote{$\dag${ Equal Contribution}  $^{*}$ Corresponding Author}
\end{abstract}    
\section{Introduction}
\label{sec:intro}

Medical image segmentation is a critical and challenging task in computer-aided medical diagnosis. By providing doctors with objective and precise references for regions of interest, well-designed medical image segmentation methods can significantly enhance the accuracy of clinical diagnosis. 
However, these performance enhancements come at the cost of increased model size and inference latency. In real-world medical applications, such as real-time detection and segmentation, there is a demand for timely execution of visual recognition tasks on resource-constrained mobile devices.
In the field of medical imaging, due to limitations of imaging principles and its specific characteristics, U-Net~\cite{unet} has become the first choice. This architectural paradigm and its variants have attained tremendous success across a wide range of medical images, including Ultrasound ~\cite{cmunet}, CT~\cite{transunet, swinunet}, Dermoscopy~\cite{egeunet}, and others~\cite{nnunet, unet++, unext}. 

\begin{figure}[t]
  \centering
   \includegraphics[width=0.9\linewidth]{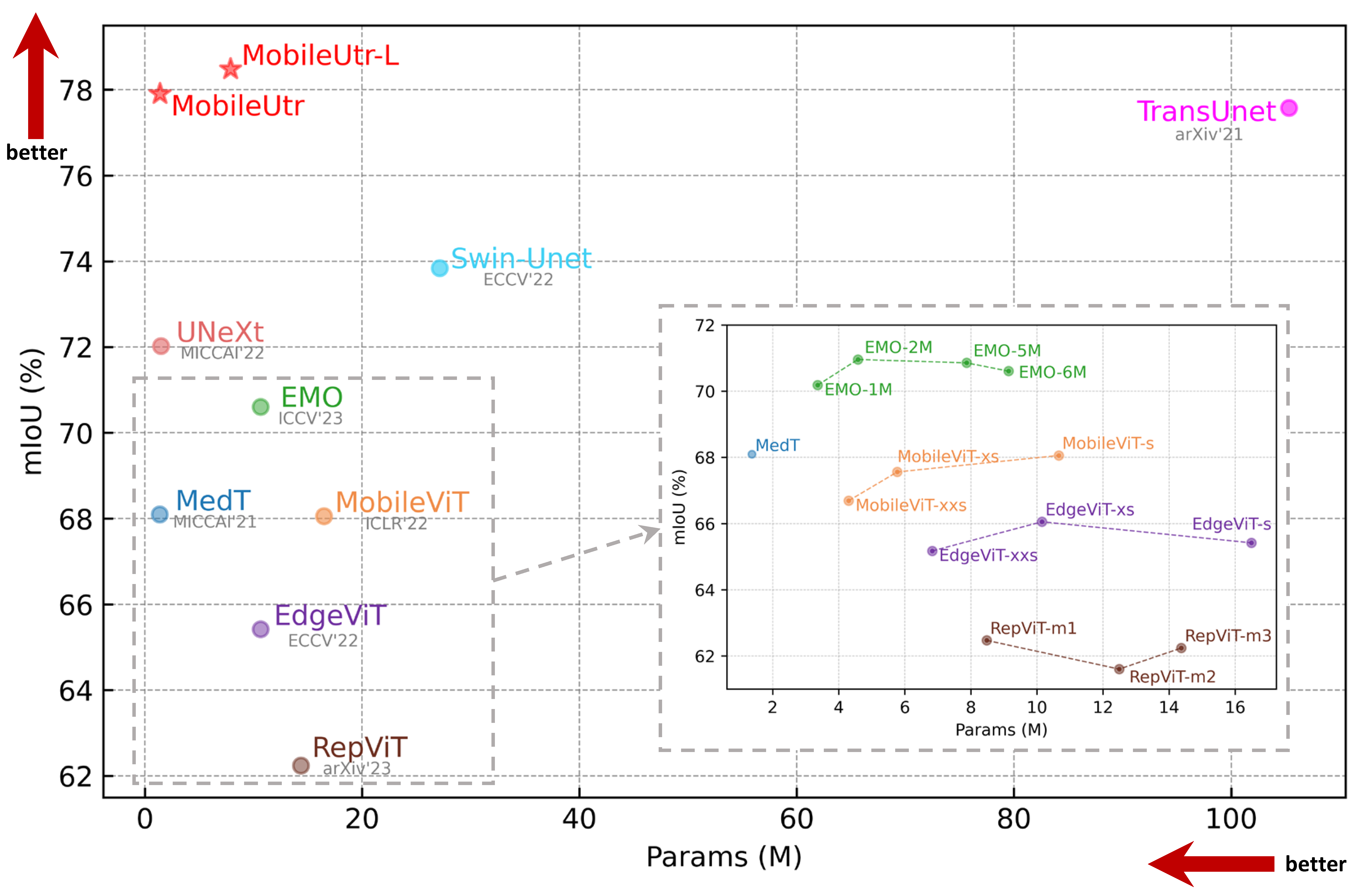}
   \caption{mIoU and Params (M) of concurrent light-weight and Transformer based methods by averaging the segmentation performance (mIoU) on the five public medical datasets.}
   \label{fig:fig1}
\end{figure}

With a recent increasing demand for storage/computing constrained applications in the medical field, mobile models with fewer parameters and lower FLOPs have attracted significant attention from researchers~\cite{mobilenet, mobilenetv2, mobilevit, emo}. In efficient model design, CNN is a cheap way to implement light-weight backbones due to \textbf{high inference efficiency} and \textbf{strong inductive bias}, and has made a great progress in medical segmentation~\cite{unext, egeunet}. However, it is inevitable that due to the local limitations of CNNs, pure CNN models cannot achieve further breakthroughs in segmentation performance~\cite{limitations}.

Compared with CNN-based methods, Transformers~\cite{transfomer} have \textbf{upper performance limitation} and not only have demonstrated robust capabilities for extracting global context information, but also have shown remarkable transferability to downstream tasks when pretrained on large-scale datasets~\cite{bert, DINO}. In the field of computer vision, this concept has evolved into the Vision Transformer (ViT)~\cite{vits}. Many studies based on this architecture have achieved significant improvements over CNNs~\cite{swintransfomer, DINO, transbts, transunet, swinunet}. To improve their performance, there has been a general trend of increasing the number of parameters in ViT~\cite{trdis, levit, cvt, DUAL-VIT, REPLACESOFTMAX}.  
However, almost ViT networks powered by \textbf{computation-hungry} self-attention consume large computation resources and cannot meet the requirements of real-time segmentation. 
Therefore, a key question comes out: \textit{how to effectively blend the computational efficiency inherent in CNN with the superior representational capacity exhibited by ViT?}

To answer it, researchers explore the fusion of CNN and Transformer models. The hybrid architecture can make the advantages of the inductive bias of CNN and the learning global context information ability of ViT to achieve better performance on medical images~\cite{transunet, transfuse, transbts}. Remarkably, both TransUnet~\cite{transunet} and TransBTS~\cite{transbts} retain the CNN structure in the encoder portion while incorporating ViT components at the bottom, yielding remarkable success in general medical segmentation tasks. 
However, while these methods absorb the performance advantages of the transformer, they do not get rid of its computational disadvantages. They still rely heavily on substantial computational resources, making them unsuitable for deployment in real-world clinical settings. 

To preserve the high performance of ViT and the high computational efficiency of CNN, we have noticed that there are several aspects that need to be carefully considered: 
1) The existence of noise, low resolution and blurred boundaries between semantics in medical images makes it difficult for Transformers to learn long-range representation between medical image patches. Inevitably, short-range relationships are further damaged, making learning more difficult; 
2) The inductive bias inherent in CNN allows to efficiently learn representations from scarce medical data using relatively few parameters, which ViT does not possess. CNN can transform the input from a pixel-level space into a latent semantic space that ViT can understand with fewer computing better.

Based on the above motivations, we revisit the relationship between CNNs and Transformers in medical image segmentation networks, dedicating to integrate the advantages of both at the infrastructure design level. For taking full advantage of the inductive bias in CNN, we try to introduce the whole design idea of Transformer into CNN. While observing that the depthwise convolution and the pointwise convolutions with inverted bottleneck exhibit similar structural similarities to MHSA and FFN in Transformers, we inductively abstract a CNN module (ConvUtr, see Section~\ref{sec:ConvUtr}) with the Transformer-like design to provide easy-learned embeddings. In addition, in order to achieve a smooth transition from the local features extracted by CNN and the global features extracted by the Transformer, we introduce the 
adaptive lightweight local-global-local (LGL) module (see Section~\ref{sec:lgl}) between CNN and Transformer to enable exchange between local and global information flows. Finally, we meticulously analyse and construct a u-shaped network for medical image segmentation, and we call this novel ViT-based lightweight network MobileUtr. To the best of our knowledge, MobileUtr represents the first, most lightweight and efficient universal medical segmentation network (e.g. 1\% higer than heavy-weight TransUnet~\cite{transunet} and 6\% higher than light-weight UNeXt~\cite{unext} in Fig.~\ref{fig:fig1}). The contributions of this paper are as follows:

\begin{enumerate}
    \item 
    We propose a Transformer-like CNN module (ConvUtr) as the patch embedding for Transformer. ConvUtr efficiently compresses medical images from the pixel space to the latent space, while providing Transformer with easy-to-understand semantic encoding.

    \item 
    We improve the adaptive Local-Global-Local (LGL) transformation as an adapter between CNN and Transformer with larger aggregation receptive field to achieve efficient exchange between local and global information flows, thereby enhancing the ability to effectively capture global context information for the transformer.

    \item 
    We validate our network on three modalities, including five different public medical datasets. Through comprehensive experimental results, the results prove that our MobileUtr can gain superior performance over the recent state-of-the-art (SOTA) approaches.
    
\end{enumerate}

\section{RelatedWork}
\label{sec:RelatedWork}

\begin{figure*}
    \centering
    \includegraphics[width = 1.0 \linewidth]{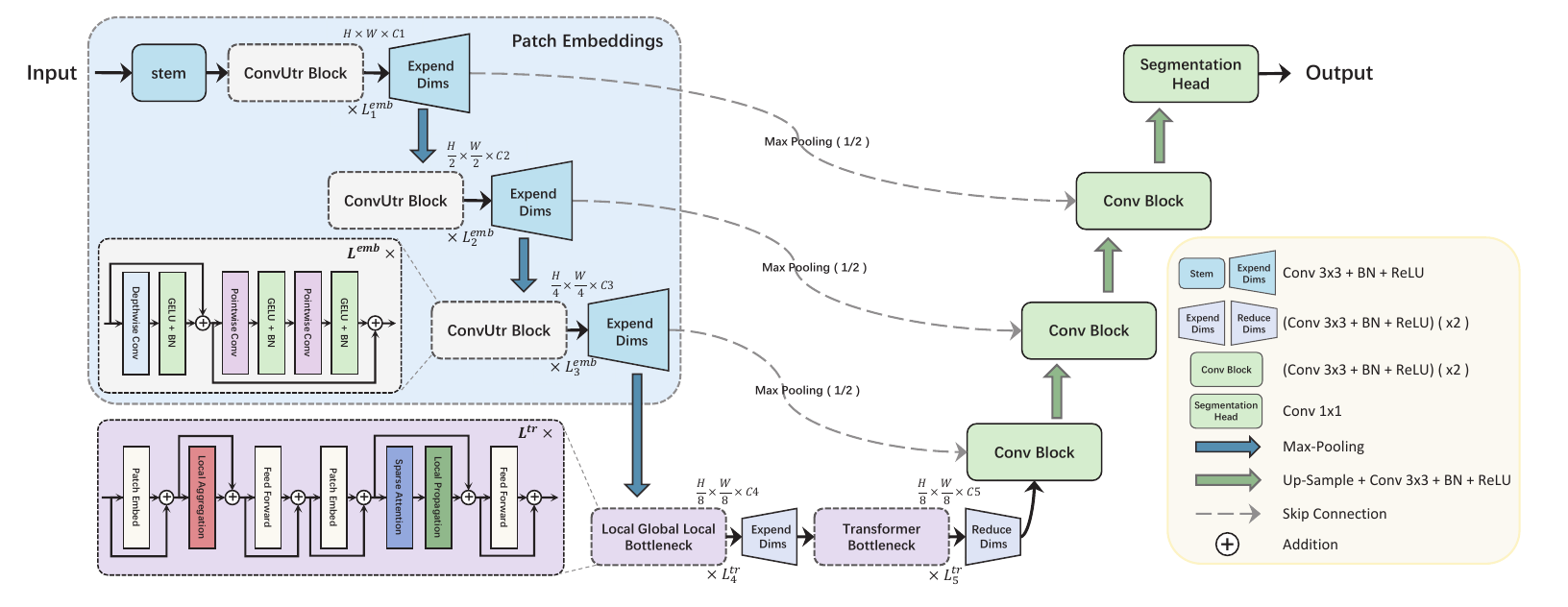}
    \caption{The architecture of MobileUtr. The encoder is divided into 5 layers. The first three layers are ConvUtr block with CNN structure, the fourth layer is adaptive LGL bottleneck, and the fifth layer is Transformer bottleneck. The first three layers of ConvUtr block mainly replace the patch embedding structure in the ViTs. The decoder is divided into 4 layers, each layer combines upsample block and convolution block with skip-connection for feature fusion.}
    \label{fig:MobileUtr}
\end{figure*}

\subsection{Light-weight networks}
\label{sec:lwn}

In early efficient model design, light-weight models based on CNN have made great progress~\cite{mobilenet, mobilenetv2, efficientnet, efficientnetv2}. It is worth noting that MobileNetV2~\cite{mobilenetv2} proposes an efficient network based on depthwise convolution combined with inverted bottleneck design, which is considered as the core design idea of efficient networks. In addition, UNeXt~\cite{unext} and EGE-Unet~\cite{egeunet} have enriched the choices in the medical vision field.

In recent years, researchers have endeavored to introduce ViTs into natural vision tasks~\cite{mobilevit, edgevits, emo, repvit}. In MobileViT~\cite{mobilevit}, the MobileNet~\cite{mobilenetv2} is integrated with ViT structure, which achieves significant success across various natural vision tasks. 
Then, EdgeViT~\cite{edgevits} proposes the local-global-local bottleneck to reduce the network parameter size. Moreover, the RepViT~\cite{repvit} and EMO~\cite{emo} propose recently further improve the performance of networks in introducing Transformers design into CNN while maintaining the light-weight. 

\subsection{Hybrid architecture of CNN and Transformer}
\label{sec:hybrid}

Because the reason that Transformers require a large number of parameters to gain the effective inductive bias capabilities, many lightweight ViT networks usually combine with CNN in natural vision tasks~\cite{mobilevit, edgevits, emo}. In medical field, the application of ViT networks has brought great difficulties due to the scarcity and limitations of medical images. Recently, researchers mix CNN with ViT and try to make up for this shortcoming~\cite{transfuse, transbts, transunet, swinunet}. Among them, Transfuse~\cite{transfuse} combines CNN and Transformer with a parallel style. TransUnet~\cite{transunet} retains the CNN in the encoder top part and the ViT at the bottom. Swin-Unet~\cite{swinunet} integrates the Swin transformer into u-shaped structure. However, most of the above methods target a single modality and obtain great segmentation performance through a large number of parameters. So far, there are few works that maintain the performance of the CNN and Transformer hybrid architecture while reducing the amount of model parameters and calculations to match the resource constraints of mobile devices.

\section{Method}
\label{sec:Method}
Constructing an effective CNN-Transformer fusion network in a lightweight manner  requires careful consideration of two key aspects: 1) Achieving a balance between the proportions of CNN and Transformer within the network. 2) Considering the distinct semantic feature requirements of CNN and Transformer, a semantic feature transformation for each layer is needed to facilitate a smooth transition.

To tackle these considerations, we develop MobileUtr, a lightweight and mobile-friendly universal medical segmentation model that combines CNN and ViT. 

\subsection{Overview of Network Architecture}
\label{sec:Overview}
The overall architecture of the proposed MobileUtr is shown in Figure~\ref{fig:MobileUtr}. MobileUtr follows 
 u-shape architecture. The encoder comprises the ConvUtr as patch embeddings, adaptive LGL bottleneck and Transformer bottleneck. The decoder consists of the progressive cascade upsampling block and convolution block for skip-connection.

\begin{table}[t]
    \renewcommand{\arraystretch}{1.2} 
    \centering
    \resizebox{1\linewidth}{!}{
        \centering
        \begin{tabular}{c|ccccc|ccccc|ccc}
        \hline
        \multirow{2}{*}{Networks} & \multicolumn{5}{c|}{Number of channels} & \multicolumn{5}{c|}{Length of blocks} & \multicolumn{3}{c}{Kernel size}\\
        \cline{2-14}
        \multicolumn{1}{c|}{}  & C1 &  C2 & C3 & C4 & C5 & $L_1^{emb}$ & $L_2^{emb}$ & $L_3^{emb}$ & $L_4^{tr}$ & $L_5^{tr}$ & K1 & K2 & K3 \\
        \hline
        \multicolumn{1}{c|}{MobileUtr}  & 16 &  32 &  64  & 64 & 128 & 1 & 1 & 3 & 3 & 3 & 3 & 3 & 7 \\
        \hline
        \multicolumn{1}{c|}{MobileUtr-L}  & 32 &  64 & 128 & 128 & 256 & 1 & 1 & 3 & 3 & 4 & 3 &  3 & 7 \\
        \hline
        \end{tabular}
    }
    \caption{MobileUtr variants.}
    \label{tab:variants}
\end{table}

The specific settings (MobileUtr and MobileUtr-L) of encoder now is presented in Table~\ref{tab:variants}, including length of block, kernel size and number of channel. The following section provides a detailed overview of the MobileUtr.

\subsection{Encoder}
\noindent{\textbf{{ConvUtr as Patch Embeddings:}}
\label{sec:ConvUtr}
Current methods use heavy-weight CNN to extract medical semantic patches to boost performance~\cite{transbts, transunet}, but the semantic patches its provided are redundant and require heavy-weight Transformer structures to match and learn global representation. Consequently, the key challenge lies in devising a lightweight Transformer, with a particular emphasis on designing a light-weight patch embedding based on CNN. This can provide Transformer with denoised, non-redundant, and highly condensed semantic patches, and also release the pressure of large parameter requirements for the Transformers.

In order to achieve the above goals, we employ CNN to emulate the behavior of Transformers. Given an image $X \in \mathcal{R}^{H \times W \times 3}$, we attempt to utilize the proposed ConvUtr block to get embeddings $X_e$ for the ViT architecture. The precise definition of ConvUtr block is as follows:
\begin{equation}
    Y_l=BN\left(\sigma\left\{DepthwiseConv\left(X_l\right)\right\}\right)+X_l
\end{equation}
\begin{equation}
    Z_l=BN\left(\sigma\left\{PointwiseConv\left(Y_l\right)\right\}\right)
\end{equation}
\begin{equation}
    X_{l+1}=BN\left(\sigma\left\{PointwiseConv\left(Z_l\right)\right\}\right) + Y_l
\end{equation}
where $X_l$ represents the output feature map of the $l$-th layer in the ConvUtr block, $Y_l$ and $Z_l$ are the intermediate variables, $\sigma$ denotes the GELU activation function~\cite{gelu}, and $BN$ denotes batch normalization. The hidden dimension between the two pointwise convolutions are four times wider than the input dimension.

To light-weighting networks while maintaining performance, the ConvUtr block employs depthwise separable convolution to emulate patch embeddings in ViTs. This combination maintains a design structure and philosophy similar to the Transformer, involving the mixing of information in spatial and channel dimensions respectively. Specifically, within the ConvUtr block, depthwise convolution (i.e., groups equal to the channels) can extract spatial dimension information as a replacement for the multi-head self-attention (MHSA). Subsequently, we employ two inverted bottleneck pointwise convolutions (referred to as FFN) to thoroughly combine spatial and channel information. Finally, we apply a convolution operation to expend the outputs feature channel of ConvUtr. We set up three ConvUtr blocks to gain semantics patch embeddings with rich representation. And the length ($L_1^{emb}$, $L_2^{emb}$, $L_3^{emb}$), kernel size (K1, K2, K3) and channels (C1, C2, C3) of each block are show in Table~\ref{tab:variants}.

In the context of transitioning between network layers, the choice of downsampling method holds significant importance. Considering the typical characteristics of medical images, which frequently exhibit low resolution and poorly defined edges, traditional pooling operations prove to be effective for noise reduction without imposing additional computational overhead~\cite{unet, nnunet}. Consequently, we select max-pooling for downsampling, using the window size of $2\times2$ and stride of 2.

\begin{figure}[!]
  \centering
  \begin{subfigure}{0.304\linewidth}
    \centering
    \includegraphics[width=1\linewidth]{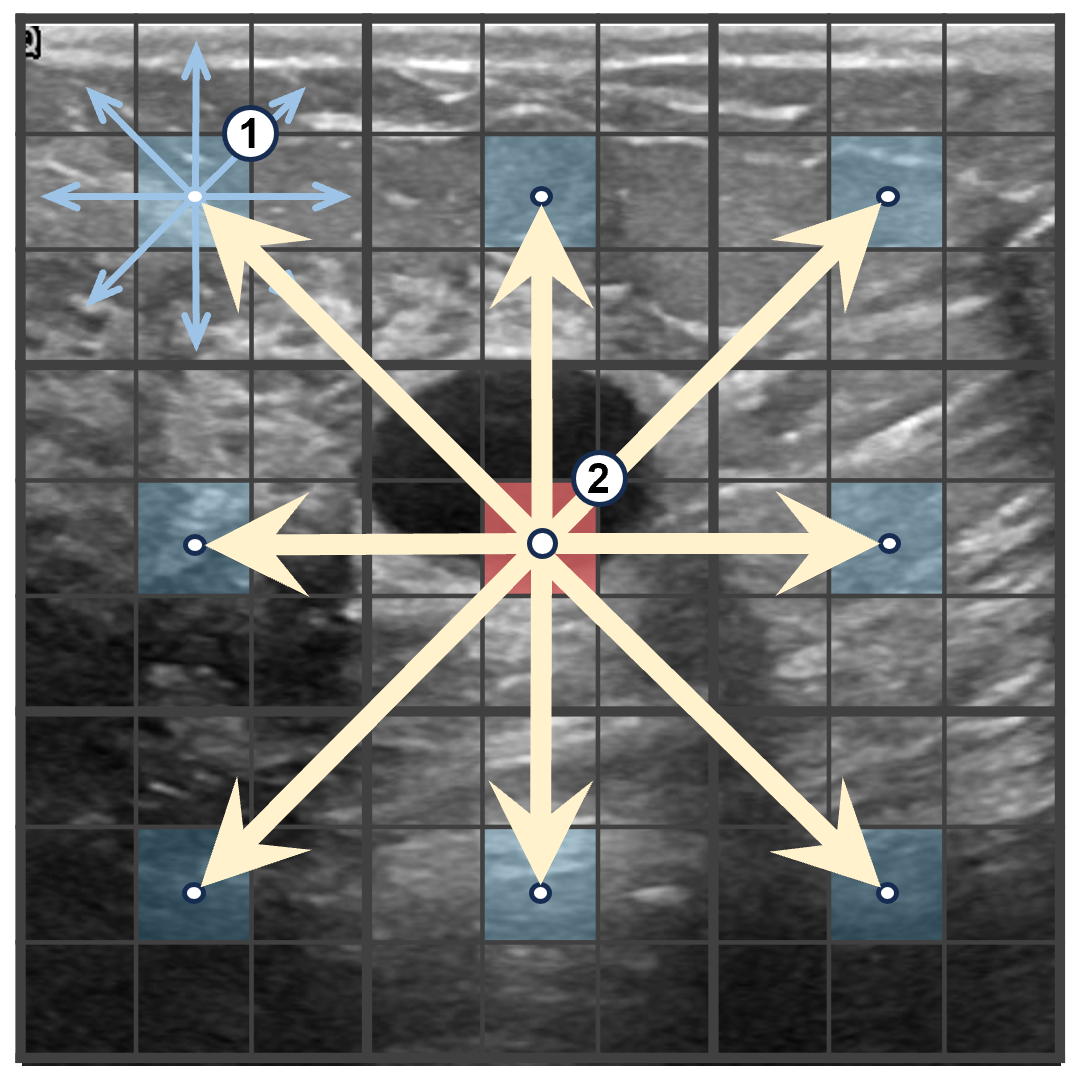}
    \caption{Transformer}
    \label{fig:Transformer}
  \end{subfigure}
  \hspace{0.1cm}
  \begin{subfigure}{0.3\linewidth}
    \centering
    \includegraphics[width=1\linewidth]{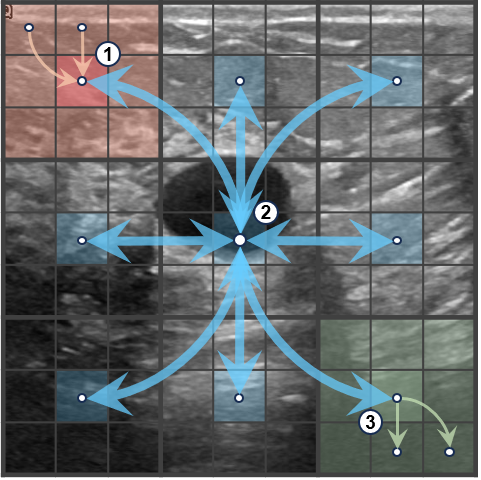}
    \caption{LGL}
    \label{fig:LGL}
  \end{subfigure}
  \hspace{0.1cm}
  \begin{subfigure}{0.3\linewidth}
    \centering
    \includegraphics[width=1\linewidth]{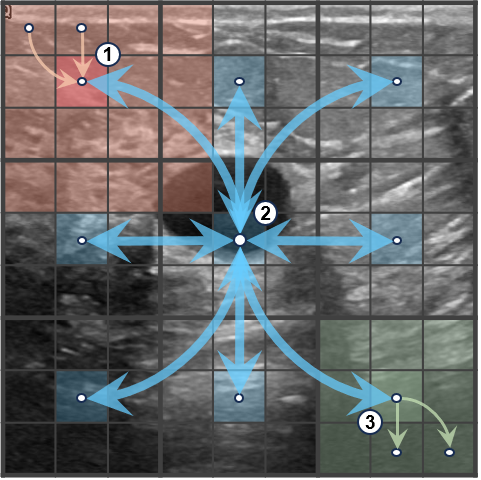}
    \caption{Adaptive LGL}
    \label{fig:Adaptive}
\end{subfigure}

   \caption{Transformer Bottleneck, LGL Bottleneck and Adaptive LGL Bottleneck. In Transformer bottleneck, every pixel sees every other pixel. In LGL and adaptive LGL bottleneck, $Local Agg (op\ 1)$, $GlobalSP (op\ 2)$, $LocalPro (op\ 3)$ are operated in sequence.}
  \label{fig:fig3}
\end{figure}

\noindent{\textbf{{Adaptive Local-Global-Local Bottleneck:}}
\label{sec:lgl}
After passing through the patch embeddings, we obtain an 8$\times$ downsampled feature map. And there is a main problem in designing a ViT bottleneck used for medical image:  \textit{When combining CNN and ViT, how can we ensure information exchange and transformation between two different structures ?}

As shown in Figure~\ref{fig:fig3}(a)-(b), compared with Transformer, LGL bottleneck \cite{edgevits} gives a reasonable structure which consisted of three operation: Local Aggregation ($LocalAgg$), Global Sparse Attention ($GlobalSP$) and Local Propagation ($LocalPro$). However, LGL bottleneck still has several disadvantages in terms of receptive field. 

$LocalPro$ of LGL bottleneck behaves akin to traditional one-dimensional window signal convolution in view of math shown in Equation~\ref{EQ:WSConv}. 
\begin{equation}
\label{EQ:WSConv}
    F(w,t) = \int_{-\infty}^{+\infty}g(u-t)f(u)e^{2\pi jwu }du
\end{equation}
where signal $F$, time $t$, frequency $w$, window time shift $u-t$.
And the key point is window size $g(u-t)$ which is similar to the kernel size ($K$) of convolution. If we can judiciously control the scope of each transformation, it can to some extent alleviate the issue of information loss during the transformation process. Therefore, in order to solve this problem, we calculate the size of the convolution kernel in advance as a prior. And we called it as adaptive LGL, which can cover the area of interest in segmentation with a larger receptive field, achieving more efficient information exchange (making red area of Figure~\ref{fig:fig3}(c) more adaptive with semantic of foreground and background).

Before reaching the ViT layer, the input undergoes a series of downsampling operations over $n$ layers. This implies that each pixel's receptive field at the ViT layer is $2^{n}$. The kernel size $K$ can be calculated as:
\begin{equation}
    K = \frac{\bar{D}}{2^{n+1}}
\end{equation}
In this context, $\bar{D}$ denotes the average diameter of segmented regions in dataset $D$. We have adjusted various aggregation scales within the local aggregation module to explore the optimal segmentation receptive field. This fine-tuning, combined with the $LocalPro$, ensures that information can be exchanged within the ViT module, maximizing its utilization and enabling effective long-range communication of information extracted in the CNN layers.

Finally, as shown in Figure~\ref{fig:fig3}(a),  we use the Transformer bottleneck as the final layer of encoder to obtain global context. The length ($L_4^{tr}$, $L_5^{tr}$), kernel size (K4, K5) and channels (C4, C5) of LGL and Transformer bottleneck are show in Table~\ref{tab:variants}. 

\subsection{Decoder with Skip-connection}

\noindent{\textbf{{Strategy for Skip-connections:}}
In the skip connection stage, achieving an appropriate fusion of global and local semantic features holds the potential to enhance segmentation performance. However, in a hybrid architecture combining CNN and Transformer, the low-level features extracted by CNN often suffer from noise interference and exhibit a significant semantics disparity compared to the high-level features of Transformer. If concatenating low-level features with decoder directly, these discrepancy hampers the overall improvement in segmentation performance.

In order to alleviate these issues from decoder's perspective, we utilize downsampling operations to the encoding features at each level of the skip connection. This operation not only eliminates additional noise interference but also ensures an appropriate receptive field during the skip-connection process, which in turn facilitates improved alignment of global and local information. Moreover, to achieve comprehensive feature fusion, we employ two convolution operations (with a kernel size of 3, stride of 1, and padding of 1) and apply ReLU activation function and batch normalization layer after each convolution.

\noindent{\textbf{{Progressive Cascade Upsampling:}}
As depicted in Figure.~\ref{fig:MobileUtr}, to effectively distinguish and capture subtle differences in medical images semantic information, we adopt a progressive cascade upsampling approach. This approach consists of multiple stages, each comprising a 2$\times$ upsampling layer, a convolution layer, a batch normalization layer, and a ReLU activation function. For the upsampling process, we utilize bilinear interpolation, which helps preserve the finer details during the upsampling operation. The convolution layer within each stage employs a kernel size of $3\times3$, a stride of 1, and padding of 1 to capture spatial dependencies and enhance feature representation.

\section{Experiment}
\label{sec:Experiment}

\begin{figure}[!]
  \centering
  \begin{subfigure}{0.19\linewidth}
    \centering
    \includegraphics[width=1 \textwidth]{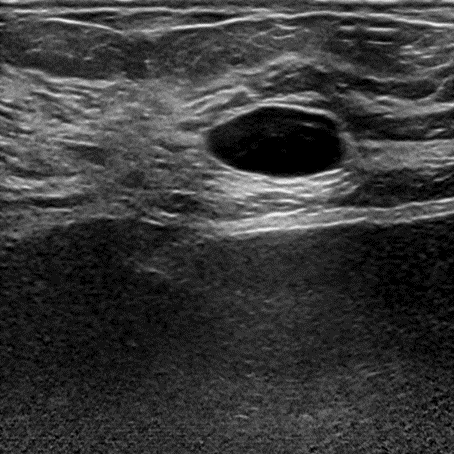}
    \caption{}
  \end{subfigure}
  \begin{subfigure}{0.19\linewidth}
    \centering
    \includegraphics[width=1 \textwidth]{sec/images/result/dataset/busi.png}
    \caption{}
  \end{subfigure}
  \begin{subfigure}{0.19\linewidth}
    \centering
    \includegraphics[width=1 \textwidth]{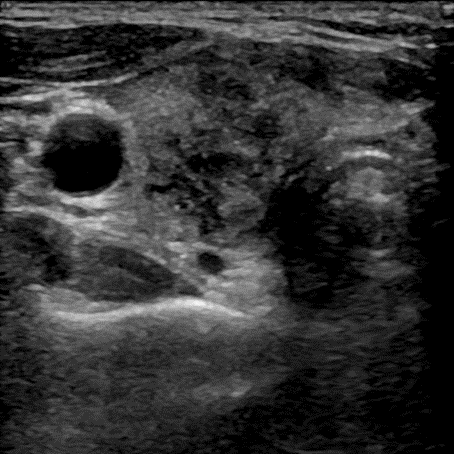}
    \caption{}
  \end{subfigure}
  \begin{subfigure}{0.19\linewidth}
    \centering
    \includegraphics[width=1 \textwidth]{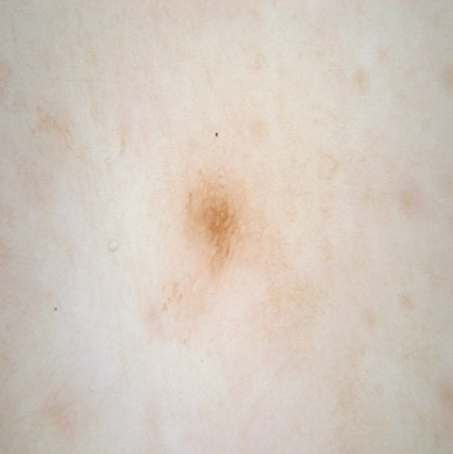}
    \caption{}
  \end{subfigure}
  \begin{subfigure}{0.19\linewidth}
    \centering
    \includegraphics[width=1 \textwidth]{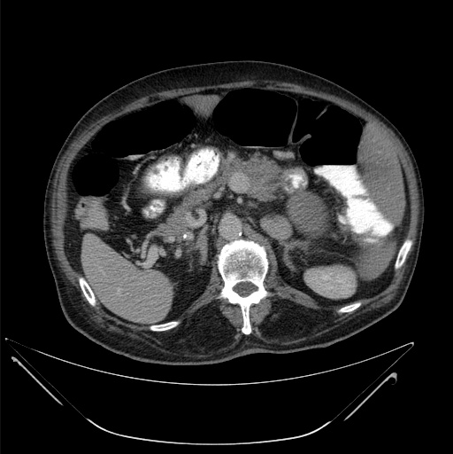}
    \caption{}
  \end{subfigure}
    \caption{Different images from (a) BUS, (b) BUSI, (c) TNSCUI, (d) ISIC 2018, (e) Synapse. These images essentially lack meaningful internal information, and the remaining valid information is also challenging to distinguish from external clutter. When the network makes judgments, we believe that it relies heavily on the distribution, and this distribution requires a larger window to capture.}
  \label{fig:MedicialImage}
\end{figure}

\subsection{Experiment Setting}

\noindent{\textbf{Dataset:}}
We select five public datasets to evaluate our network as well as other state-of-the-art networks. The dataset used in this study comprises three main modalities: CT (Synapse~\cite{DATA_SYNAPSE} with 30 cases), Ultrasound (BUS~\cite{DATA_BUS} with 562 images, BUSI~\cite{DATA_BUSI} with 647 images, TNSCUI~\cite{DATA_TNSCUI} with 4554 images), and Dermoscopy images (ISIC 2018~\cite{DATA_ISIC2018} with 2594 images). We use a 70/30 split on the four medical image datasets (BUS, BUSI, TNSCUI, ISIC2018) for training and validation thrice. In addition, we randomly split Synapse dataset into 18 cases for training (2212 axial slices) and 12 cases for validation.

\renewcommand{\multirowsetup}{\centering}  
\begin{table*}[!]
\resizebox{1\linewidth}{!}
{
\begin{tabular}{r | r | r | r | c c | c c | c c | c c}
\hline 

\multirow{3}{*}{Network} & \multirow{3}{*}{Params↓} & \multirow{3}{*}{FPS↑} & \multirow{3}{*}{GFLOPs↓} & \multicolumn{6}{c|}{Ultrasound} &  \multicolumn{2}{c}{Dermoscopy} \\
\cline{5-12}
&&&& \multicolumn{2}{c|}{BUS  (\%)} & \multicolumn{2}{c|}{BUSI  (\%)} & \multicolumn{2}{c|}{TNSCUI  (\%)} & \multicolumn{2}{c}{ISIC  (\%)} \\
\cline{5-12}
&&&& IOU &F1 & IOU &F1 & IOU &F1 & IOU &F1\\
\hline
\textit{{\color{Gray} heavy-weight medical network}} &&&&&&&&& \\

\rowcolor{mygray} U-Net~\cite{unet}   & 34.52\,M & 139.32 & 65.52 & 86.73±1.41 & 92.46±1.17 & 68.61±2.86 & 76.97±3.10 
                                 & 75.88±0.18 & 84.24±0.07 & 82.18±0.87 & 89.97±0.52 \\
\rowcolor{mygray} CMU-Net~\cite{cmunet} & 49.93\,M & 93.19  & 91.25 & 87.18±0.59 & 92.89±0.41 & 71.42±2.65 & 79.49±2.92 
                                 & 77.12±0.49 & 85.35±0.50 & 82.16±1.06 & 89.92±0.62\\
\rowcolor{mygray} nnUNet~\cite{nnunet}  & 26.10\,M   & ---    & 12.67   & \underline{87.51±1.01} & \underline{93.02±0.73} & 72.11±3.51 & 80.09±3.77 & \textbf{78.99±0.14} & \textbf{86.85±0.15} & \underline{83.31±0.59} & 89.84±0.50\\
\rowcolor{myellow} Swin-Unet~\cite{swinunet} & 27.14\,M  & 392.21 & 5.91  & 85.27±1.24 & 91.99±0.75 & 63.59±4.96 & 76.94±4.12 
                                    & 75.77±1.29 & 85.82±0.91 & 82.15±1.44 & 89.98±0.87\\ 
\rowcolor{myellow} TransUnet~\cite{transunet} & 105.32\,M & 112.95 & 38.52 & 87.35±1.24 & 92.88±0.88 & 71.39±2.37 & 79.85±2.59 
                                    & 77.63±0.14 & 85.76±0.20 & 83.17±1.25 & \textbf{90.57±0.72}\\           
\hline                                    
\textit{{\color{Gray} light-weight natural network}} &&&&&&&&& \\

\rowcolor{myellow} MobileViT-s~\cite{mobilevit} & 16.49\,M  & 243.60 & 2.71  & 82.57±1.38 & 89.99±1.17 & 64.28±3.78 & 74.68±3.81 
                                    & 71.64±0.28 & 81.60±0.25 & 80.12±0.42 & 87.89±0.43\\
\rowcolor{myellow} EdgeViT-s~\cite{edgevits} & 10.66\,M  & 291.44 & 2.13  & 81.32±1.23 & 89.13±0.99 & 61.12±3.69 & 71.79±4.00 
                                    & 68.74±0.29 & 79.19±0.36 & 79.06±0.56 & 87.06±0.55\\
\rowcolor{myellow} RepViT-m3~\cite{repvit}  & 14.37\,M  & 238.90 & 2.79   & 76.51±1.38 & 85.79±1.12 & 56.07±3.51 & 67.62±3.98 & 66.21±0.66 & 77.09±0.49 & 78.34±0.61 & 86.62±0.48\\
\rowcolor{myellow} EMO-6m~\cite{edgevits} & 10.66\,M  & 291.44 & 2.13  & 84.89±0.86 & 91.58±0.63 & 67.50±4.26 &77.71±4.23 & 74.02±0.39 & 83.53±0.25 & 81.31±0.73 & 88.74±0.57\\  

\textit{{\color{Gray} light-weight medical network}} &&&&&&&&& \\

\rowcolor{mygray} UNeXt~\cite{unext}   & 1.47\,M   & 650.48 & 0.58  & 84.73±1.23 & 91.20±0.94 & 65.04±2.71 & 74.16±2.84
                                    & 71.04±0.17 & 80.46±0.16 & 82.10±0.88 & 89.93±0.46\\
\rowcolor{mygray} EGE-Unet~\cite{egeunet}  & 0.072\,M  & 303.08 & 0.045 & 84.72±1.28 & 91.72±0.75 & 58.90±2.97 & 74.11±2.34
                                    & 74.47±0.43 & 85.36±0.28 & 82.19±1.31 & \underline{90.22±0.79}\\
\rowcolor{myellow} MedT~\cite{medt}    & 1.37\,M   & 22.97  & 2.40  & 80.81±2.77 & 88.78±1.96 & 63.36±1.56 & 73.37±1.63 
                                    & 71.00±2.68 & 80.87±2.16 & 81.79±0.94 & 89.74±0.53\\
\hline
\rowcolor{myellow} \textbf{MobileUtr}  & 1.39\,M   & 326.24 & 2.51  & 87.28±0.83 & 92.90±0.63 & \underline{72.88±2.72} & \underline{81.18±3.05}
                                       & 77.70±0.50 & 85.90±0.41 & 83.23±0.39 & 89.86±0.28\\
\rowcolor{myellow} \textbf{MobileUtr-L} & 7.88\,M  & 279.42  & 3.70 & \textbf{87.63±0.91} & \textbf{93.13±0.61} & \textbf{73.91±2.65} & \textbf{82.16±2.64}
                                       & \underline{78.24±0.38} & \underline{86.37±0.28}  & \textbf{83.31±0.36} & 89.88±0.25\\
\hline
\end{tabular}
}
\caption{Result on Ultrasound and Dermoscopy Datasets.  \textbf{val} (bold) / \underline{val} (underline) : top method / second method. White and \colorbox{myellow}{gray} are backgrounds indicate CNN-based and Transformer-based.  \label{tab:ExOnUltrasound}}
\end{table*}

\noindent{\textbf{{Evaluation Metrics and Comparison Methods:}}
In this study, we primarily utilize widely used evaluation metrics, including Intersection over Union  (IoU) and F1 score for the BUS, BUSI, TNSCUI and ISIC 2018 datasets; Hausdorff Distance (HD95), Dice and mIoU for the Synapse dataset.
To evaluate the performance of medical image segmentation, We selected 12 popular medical segmentation models, including heavy-weight medical image networks: U-Net~\cite{unet}, CMU-Net~\cite{cmunet}, nnUNet~\cite{nnunet} TransUnet~\cite{transunet}, Swin-Unet~\cite{swinunet}; light-weight natural image networks: MobileViT~\cite{mobilevit}, EdgeViT~\cite{edgevits}, RepViT~\cite{repvit}, EMO~\cite{emo}; light-weight medical image models: MedT~\cite{medt}, UNeXt~\cite{unext}, EGE-Unet~\cite{egeunet}.

\noindent{\textbf{{Implement details:}}
The loss between the predicted and ground truth is defined as a combination of binary cross entropy (BCE) and dice loss (Dice). 
We resize all training cases of five datasets to 256$\times$256 and apply random rotation and flip for simple data augmentations. In addition, we use the SGD optimizer with a weight decay of 1e-4 and a momentum of 0.9 to train the networks. The initial learning rate is set to 0.01, and the poly strategy is used to adjust the learning rate. The batch size is set to 8 and the training epochs are 300. All the experiments are conducted using a single NVIDIA GeForce RTX4090 GPU.

\subsection{Analysis of Experimental Results on Images}
\subsubsection{Experiments on Ultrasound Images}
In Figure~\ref{fig:MedicialImage}, we present illustrative results showcasing the performance of various algorithms. The visual representation clearly demonstrates that our proposed MobileUtr surpasses other SOTA algorithms in terms of visual quality. To ensure a robust evaluation, the subsequent sections will provide an in-depth analysis of quantitative results.

In ultrasound image segmentation tasks (BUS, BUSI, TNSCUI datasets). Our proposed MobileUtr is compared to state-of-the-art methods mentioned in Table~\ref{tab:ExOnUltrasound}. The experimental results demonstrate that our MobileUtr achieves the best performance, striking a better balance between accuracy and computational cost. 

Specifically, in the BUS and BUSI dataset, nnUNet~\cite{nnunet} achieves the highest IoU and F1 scores. However, our network achieves comparable results while maintaining a significantly smaller model size (1.39 M vs 26.10 M), improved computational efficiency (2.51 GFLOPs vs 12.67 GFLOPs). Although nnUNet's performance is high, once we expand the network dimension, MobileUtr-L achieves the best performance with IoU score of 87.63 and 73.91 (0.1\% and 1.8\% higher than nnUNet). Even in largest TNSCUI dataset, MobileUtr, competitive performance can be obtained while maintaining the minimum parameters.

\renewcommand{\multirowsetup}{\centering}  
\begin{table*}[!]

\resizebox{1\linewidth}{!}
{
\begin{tabular}{r | c | c | c | c c c c c c c c   }
\hline 

\multirow{2}{*}{Network} & \multirow{2}{1cm}{mIoU↑} 
& \multirow{2}{1cm}{Dice↑} & \multirow{2}{1cm}{HD95↓}
& \multicolumn{8}{c}{Synapse (\%)} \\
\cline{5-12}
&&&& Aorta & Gallbladder & Kidney (L) & Kidney (R) & Liver & Pancreas & Spleen & Stomach \\
\hline
\textit{{\color{Gray} heavy-weight medical network}} &&&&&&& \\
\rowcolor{myellow} TransUnet~\cite{transunet} & \underline{68.33} & 79.12 & \textbf{23.76}
          & 78.47 & 41.83 & \textbf{75.26} & \textbf{72.59} & \underline{89.90} & \underline{44.87} & \underline{80.07} & \underline{63.68} \\
\rowcolor{myellow} Swin-Unet~\cite{swinunet}  & 62.42 & 74.13 & 28.54 
          & 70.43 & 35.73 & 68.58 & 61.64 & 88.52 & 37.01 & 79.48 & 57.97\\
\hline
\textit{{\color{Gray} light-weight natural network}} &&&&&&& \\
\rowcolor{myellow} MobileViT-s~\cite{mobilevit} & 41.66 & 55.57 & 30.81 
          & 24.63 & 16.58 & 50.94 & 45.85 & 77.74 & 17.90 & 58.80 & 40.83\\
\rowcolor{myellow} EdgeViT-s~\cite{edgevits}   & 36.86 & 50.68 & 31.82 
          & 21.97 & 12.23  & 41.42 & 38.96 & 76.43 & 16.00 & 50.71 & 37.16\\
\rowcolor{myellow} RepViT-m3~\cite{repvit}    & 34.07 & 47.61 & 30.69
          & 20.71 & \ \;7.16 & 33.56 & 37.28 & 74.33 & 13.28 & 48.13 & 38.09\\
\rowcolor{myellow} EMO-6m~\cite{emo}   & 45.30 & 59.47 & 27.68 
        & 28.42 & 22.85 & 51.88 & 48.19 & 79.50 & 22.19 & 60.80 & 48.54\\
\textit{{\color{Gray} light-weight medical network}} &&&&&&& \\
UNeXt~\cite{unext}     & 57.22 & 69.99 & 41.43 
          & 69.35 & 36.47 & 63.21 & 50.45 & 85.09 & 28.87 & 72.34 & 52.02\\
\rowcolor{myellow} MedT~\cite{medt}      & 43.51 & 55.21 & 60.06  
          & 67.49 & \ \;1.63  & 61.82 & 49.81 & 36.11 & 20.26 & 66.33 & 44.64\\
\hline
\rowcolor{myellow} \textbf{MobileUtr}    & 68.17 & \underline{79.13} & 30.96  
             & \textbf{79.64} & \underline{45.96} & \underline{74.93} & \underline{68.69} & \textbf{90.40} & 43.51 & \textbf{81.21} & 60.98\\
\rowcolor{myellow} \textbf{MobileUtr-L}  & \textbf{69.09} & \textbf{79.90} & \underline{26.49} 
             & \underline{78.99} & \textbf{49.14} & 72.55 & 68.29 & 88.87 & \textbf{50.10} & 79.57 & \textbf{65.19}\\
\hline
\end{tabular}
}
\caption{Result on Computed Tomograph Dataset. \textbf{val} (bold) / \underline{val} (underline) : top method / second method. White and \colorbox{myellow}{gray} are backgrounds indicate CNN-based and Transformer-based. \label{tab:ct}}
\end{table*}

\begin{figure}[!]
  \centering
  \begin{subfigure}{0.11\linewidth}
    \centering
    \includegraphics[width=1 \textwidth]{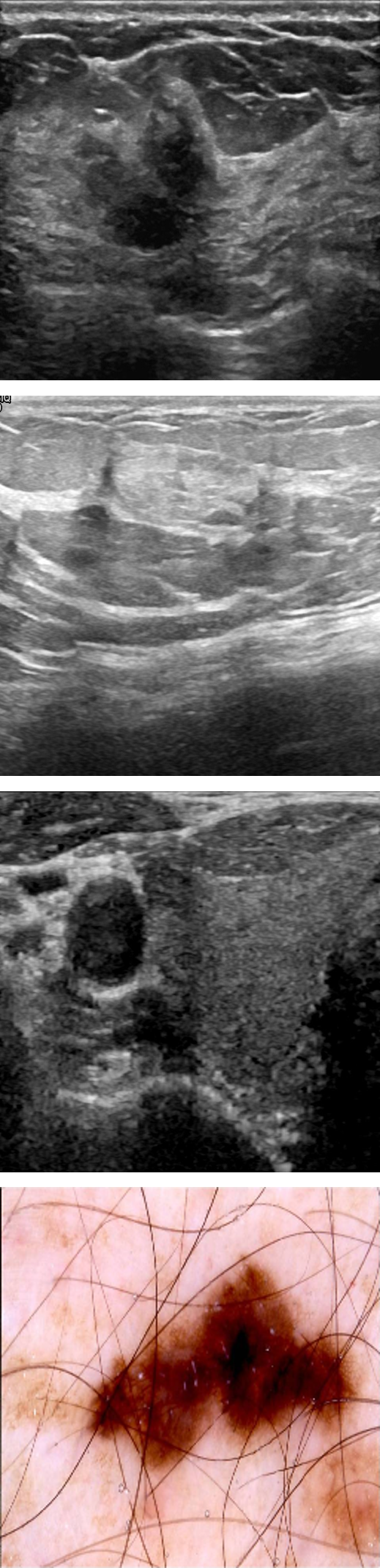}
    \caption{}
  \end{subfigure}
  \begin{subfigure}{0.11\linewidth}
    \centering
    \includegraphics[width=1 \textwidth]{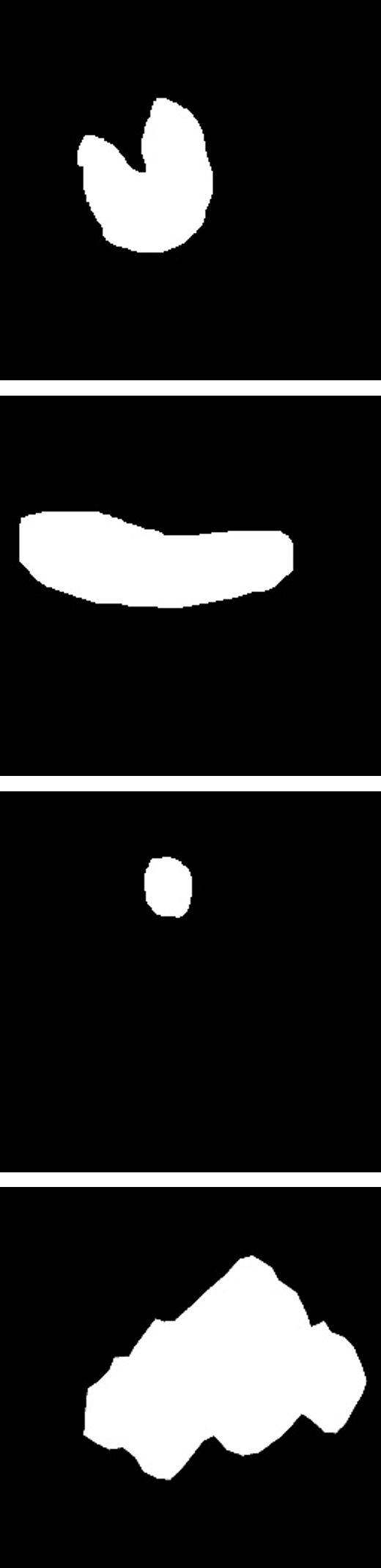}
    \caption{}
  \end{subfigure}
  \begin{subfigure}{0.11\linewidth}
    \centering
    \includegraphics[width=1 \textwidth]{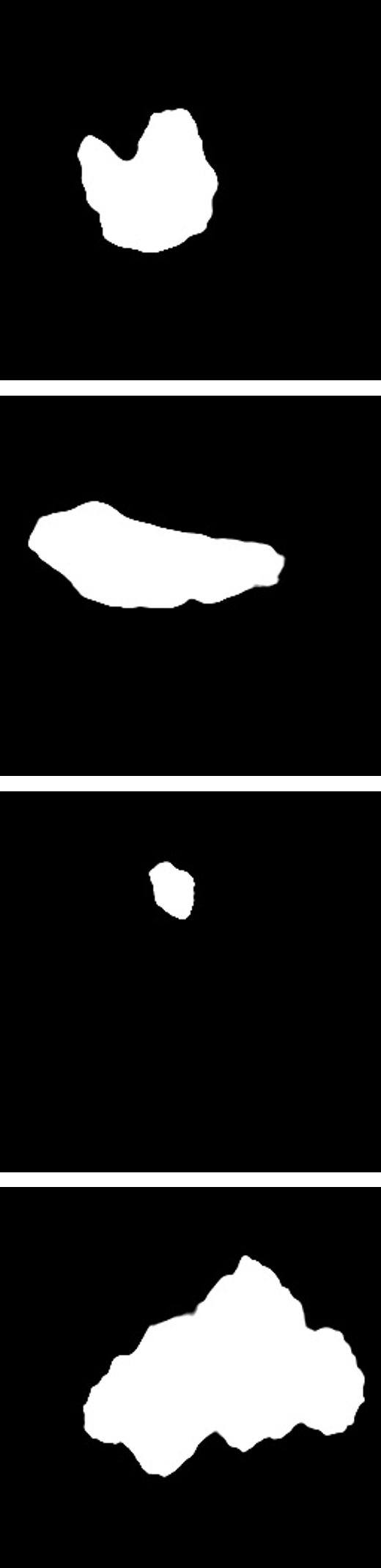}
    \caption{}
  \end{subfigure}
  \begin{subfigure}{0.11\linewidth}
    \centering
    \includegraphics[width=1 \textwidth]{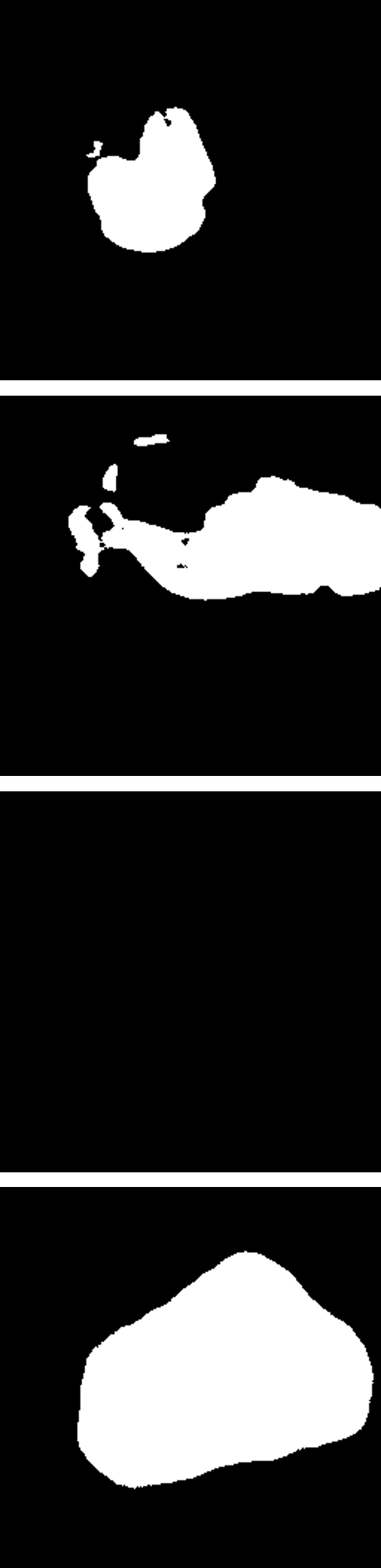}
    \caption{}
  \end{subfigure}
  \begin{subfigure}{0.11\linewidth}
    \centering
    \includegraphics[width=1 \textwidth]{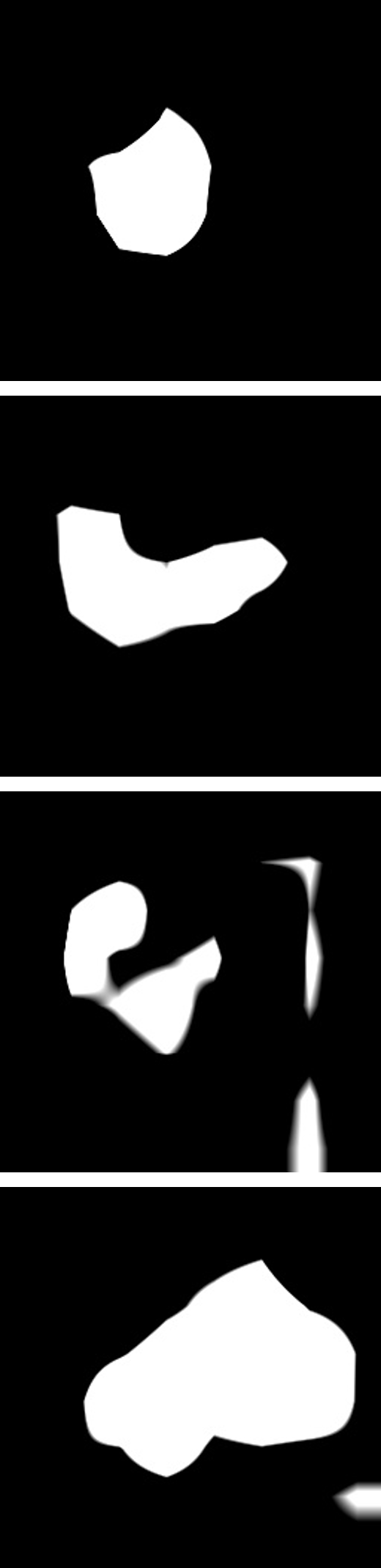}
    \caption{}
  \end{subfigure}
  \begin{subfigure}{0.11\linewidth}
    \centering
    \includegraphics[width=1 \textwidth]{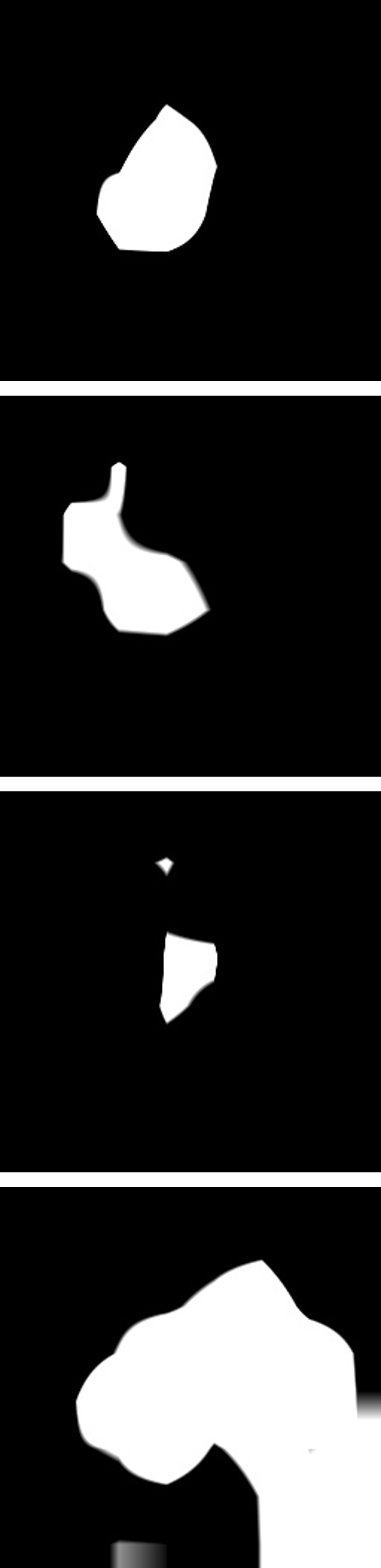}
    \caption{}
  \end{subfigure}
  \begin{subfigure}{0.11\linewidth}
    \centering
    \includegraphics[width=1 \textwidth]{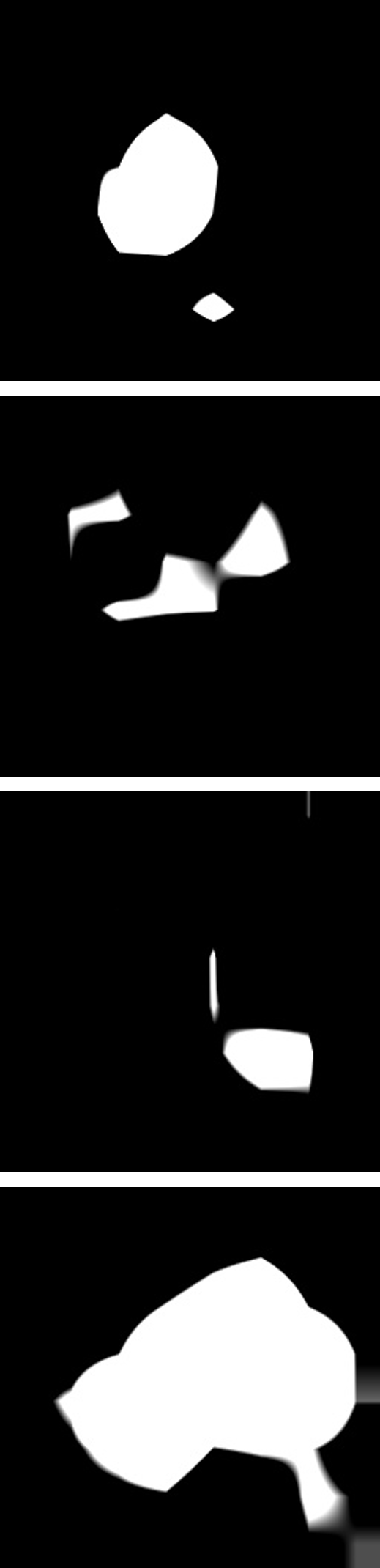}
    \caption{}
  \end{subfigure}  
  \begin{subfigure}{0.11\linewidth}
    \centering
    \includegraphics[width=1 \textwidth]{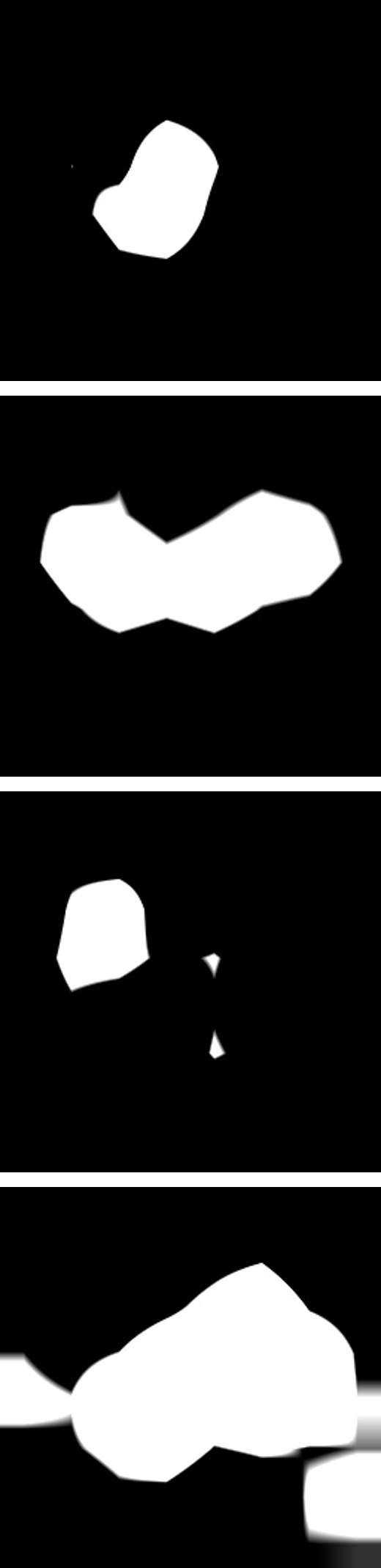}
    \caption{}
  \end{subfigure}
    \caption{Visualization Results on Ultrasound and Dermoscopy Datasets. (a) Original Medical Image (b) Ground Truth (c) MobiltUtr (d) nnUNet (e) UNeXt (f) MobileViT (g) EdgeViT (h) RepViT (i) EMO.}
  \label{fig:MedicialImage}
\end{figure}

\begin{figure*}[htbp]
    \flushright
    \begin{subfigure}{0.11\linewidth}
        \centering
        \includegraphics[width=1 \textwidth]{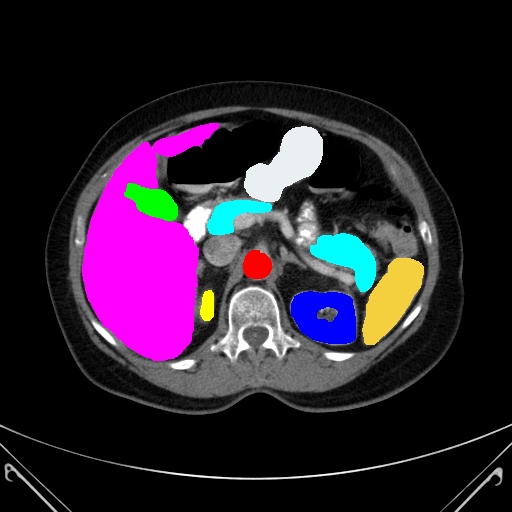}
        \caption{Ground Truth}
    \end{subfigure}
        \begin{subfigure}{0.11\linewidth}
        \centering
        \includegraphics[width=1 \textwidth]{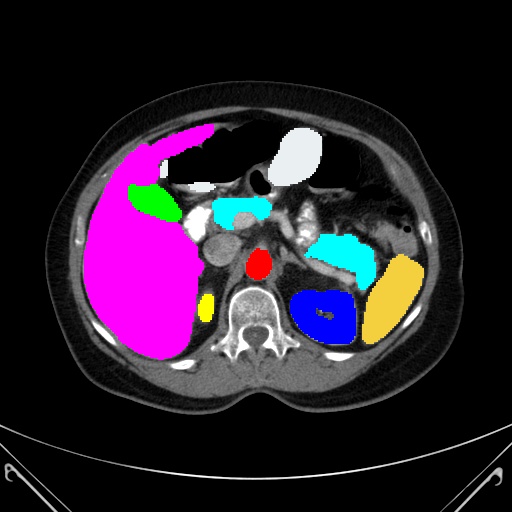}
        \caption{MobileUtr}
    \end{subfigure}
    \begin{subfigure}{0.11\linewidth}
        \centering
        \includegraphics[width=1 \textwidth]{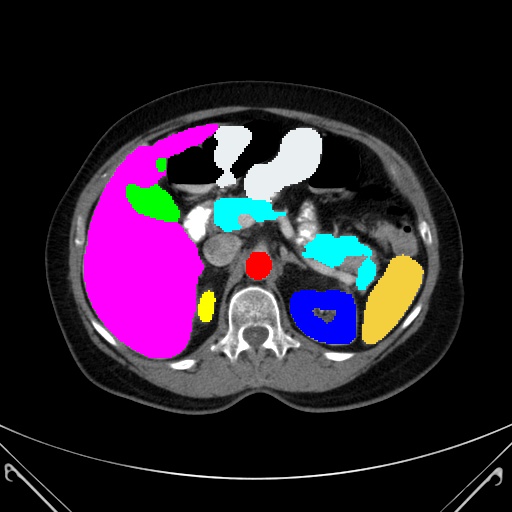}
        \caption{TransUnet}
    \end{subfigure}
    \begin{subfigure}{0.11\linewidth}
        \centering
        \includegraphics[width=1 \textwidth]{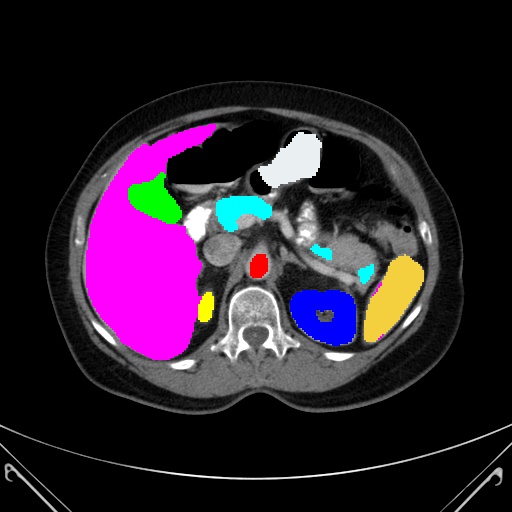}
        \caption{UNeXt}
    \end{subfigure}
    \begin{subfigure}{0.11\linewidth}
    \centering
        \includegraphics[width=1 \textwidth]{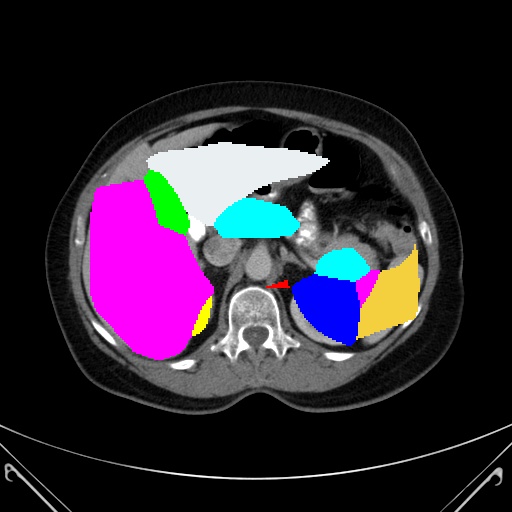}
        \caption{MobileViT}
        \end{subfigure}
    \begin{subfigure}{0.11\linewidth}
    \centering
    \includegraphics[width=1 \textwidth]{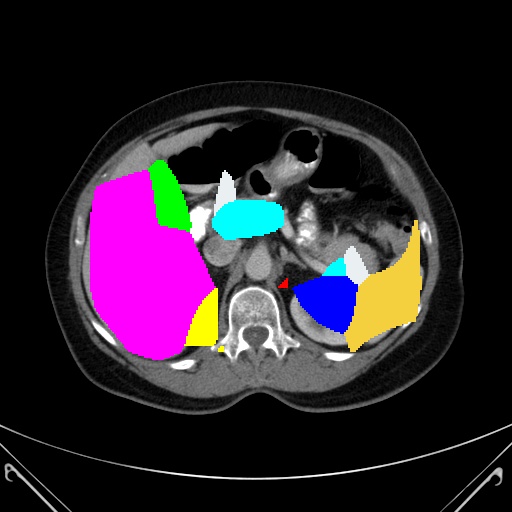}
        \caption{EdgeViT}
        \end{subfigure}
        \begin{subfigure}{0.11\linewidth}
    \centering
        \includegraphics[width=1 \textwidth]{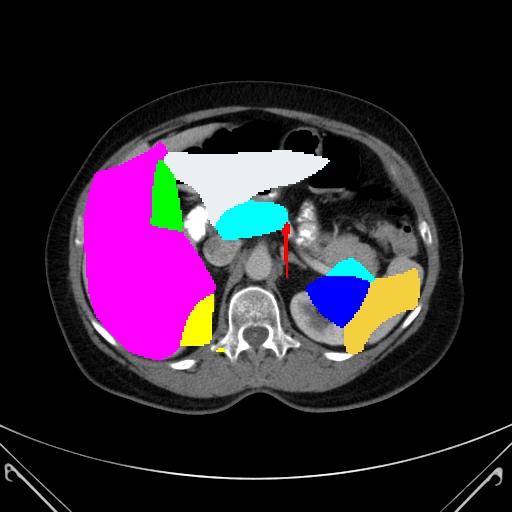}
        \caption{RepViT}
        \end{subfigure}  
    \begin{subfigure}{0.11\linewidth}
        \centering
        \includegraphics[width=1 \textwidth]{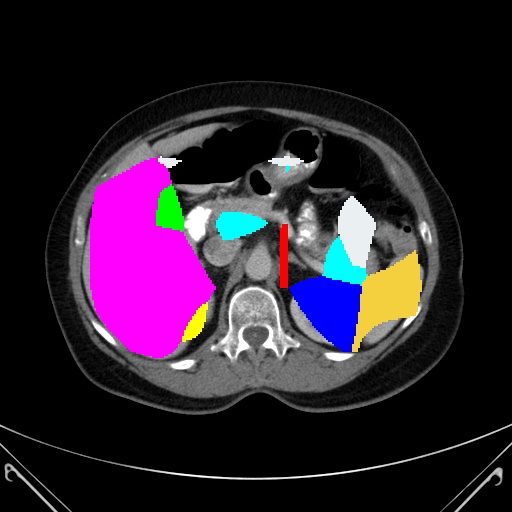}
        \caption{EMO}
    \end{subfigure}
    \begin{subfigure}{0.075\linewidth}
        \centering
        \includegraphics[width=1 \textwidth]{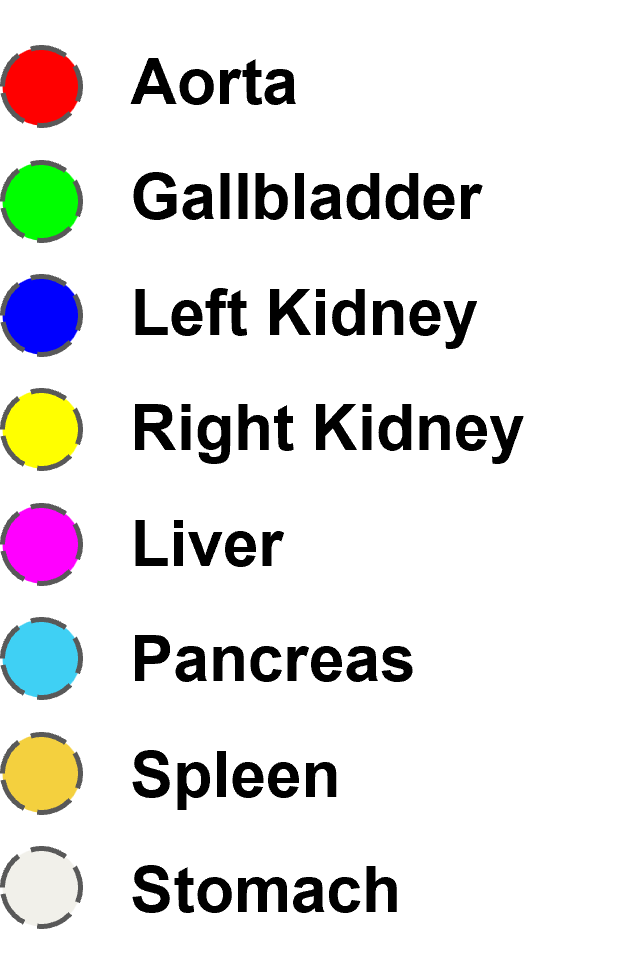}
        \vspace{0.0005cm}
    \end{subfigure}
   
    \caption{Visualization Results on Computed Tomograph Dataset.}
\label{fig:Synapse}
\end{figure*}

Moreover, the lightweight CNNs like UNeXt~\cite{unext} and EGE-Unet~\cite{egeunet} do not yield satisfactory results. While their parameter and computation requirements have significantly decreased, their corresponding performance has also declined. This phenomenon is also observed in ViT networks. When we employ networks like MobileViT~\cite{mobilevit}, EdgeViT~\cite{edgevits}, RepViT~\cite{repvit}, and EMO~\cite{emo}, their effectiveness is limited as these networks are originally designed for specific tasks in natural images. When applied to medical images, their performance is much lower than MobileUtr. On the other hand, networks that combine CNN and ViT, such as TransUnet~\cite{transunet} (105.32M parameters, 112.95 FPS, 38.52 GFLOPs) and Swin-Unet~\cite{swinunet} (27.14M parameters, 392.21 FPS, 5.91 GFLOPs), achieve certain levels of success but inevitably face a trade-off between performance and computational burden.

However, with the special designation of encoder, our MobileUtr maintains a nearly smallest light-weight model while achieving almost the best performance. And in the case of MobileUtr-L, it outperforms other models. This indicates the effectiveness and correctness of our encoder's patch embedding and combining strategy CNNs with Transformers. Furthermore, MobileUtr stands as the first successful model in accomplishing transformer light-weighting, setting a new benchmark in medical domain.

\subsubsection{Experiments on Dermoscopy Images}
In dermoscopy experiments, we focus on the challenging task of skin cancer segmentation under natural light conditions. As presented in Table~\ref{tab:ExOnUltrasound}, MobileUtr and MobileUtr-L exhibit the highest accuracy in skin cancer segmentation. In particular, EGE-Unet demonstrated a significant reduction in parameter count and computational burden without a substantial decrease in accuracy. However, this is because the dermoscopy dataset contains many details such as texture, contrast, and clear edges. Additionally, the similarity distribution between the training and the testing data greatly simplifies network training, allowing all networks to fit well.

It is worth highlighting that nnUNet and TransUnet achieve optimal performance on ultrasound image datasets (BUS, BUSI, TNSCUI) as well as the dermoscopy dataset (ISIC 2018). This is attributed to their nature as general-purpose medical segmentation networks like our MobileUtr. Indeed, the majority of medical segmentation tasks, such as ultrasound images and computed tomography (CT) images, involve weak target segmentation. As shown in Table~\ref{tab:ExOnUltrasound}, the remaining lightweight models experience significant performance drops, indicating that their attempts at lightweight design are unsuccessful. A real-working lightweight model should be effective across the majority of tasks, and this is where the value of the network proposed in this paper lies.

\subsubsection{Experiments on Computed Tomography Images}

In our comprehensive evaluation, we also included CT images, which constitute a common and important image type in medical segmentation tasks. It is essential to note that CT images inherently exist in 3D space, allowing for the application of both 3D and 2D segmentation methods. For this evaluation, we choose the 2D slice segmentation approach, which involves independently segmentation individual slices of the CT volume.

The experimental results are summarized in Table~\ref{tab:ct}. Notably, apart from the TransUnet, both lightweight transformer-based networks and lightweight CNNs-based networks experience a significant decline in performance. For instance, compared to MobileUtr, models such as MedT, EdgeViT, and MobileViT, among others, exhibit performance decreases of up to 20\%. Similarly, CNNs networks also demonstrate a performance decrease of approximately 10\%.

These findings underscore the superior performance of our proposed lightweight network MobileUtr, in CT multi organ segmentation tasks. Unlike other models, MobileUtr maintain its high performance (mIoU of 69.09\% and Dice of 79.90\%). Additionally, as shown in Figure~\ref{fig:Synapse}, our network achieves well-balanced and well-delineated segmentation across various organs in the CT images. Considering its compact size and high frame rate, this further highlights the suitability of our network for deployment on edge devices, ensuring efficient and effective medical image segmentation in real-time applications.

\subsection{Ablation Study}

\noindent{\textbf{{Ablation study on each blocks:}}
To comprehensively evaluate the proposed MobileUtr, we conduct extensive ablation experiments on the Synapse dataset to assess the contribution of each module. The ablation study results are presented in Table~\ref{tab:ablation}.

Initially, we utilize the first three layers of ResNet34\cite{RESNET} (patch embeddings) and pure ViT as encoder, omitting skip connections, which yields an mIoU of 63.76. Next, we replace pure ViT with LGL bottleneck, the network computing cost is reduced. Subsequently, we replace patch embedding with ConvUtr block while ensuring minimal computational cost. This modification result in a significant reduction in model parameters to 1.32 M, a 34$\times$ decrease in GFLOPs, and a 3$\times$ improvement in inference time, with only a slight decrease in segmentation performance. It shows that ConvUtr block successfully provides suitable encoding information to the Transformer while effectively minimizing computational costs. Then, we replace the LGL bottleneck with the Adaptive LGL bottleneck. We observe a minimum 1\% increase in mIoU, indicating the Adaptive LGL can achieve better local and global information flow exchange in medical domains.

Furthermore, we progressively incorporate additional skip connections from top to bottom. Remarkably, as the number of skip connections increase, the network's segmentation performance continue to improve while maintaining low computational costs. This observation highlights the effectiveness of skip connections in providing local detailed information to the network, thereby enhancing its knowledge transfer capability. The network achieve its highest segmentation performance of 68.17\% when three skip connections are utilized.

These ablation experiments manifest the significance of each module in MobileUtr and illuminate their respective contributions. The findings suggest that ConvUtr Block enables efficient encoding, while skip connections facilitate the integration of local details, ultimately enhancing the network's segmentation performance.

\renewcommand{\multirowsetup}{\centering}  
\begin{table}[!t]

\resizebox{1\linewidth}{!}
{
\begin{tabular}{c c| r r r | c}
\hline 

\multicolumn{2}{c|}{Network}
& \multicolumn{4}{c}{Metrics(\%)}\\
\cline{0-5}
Encoder & Decoder & Params↓  & GFLOPs↓ &  FPS↑ & mIoU↑\\
\hline
ResNet34 + ViTs  & w/o skip & 25.65 & 85.95 & 77.75 & 63.76\\
ResNet34 + LGL  & w/o skip & 22.11 & 81.41 & 100.06 & 63.25\\
ConvUtr + LGL  & w/o skip & 1.32 & 2.37 & 340.26 & 63.16\\
ConvUtr + Adaptive LGL   & w/o skip & 1.34 & 2.39 & 339.16 & 64.40\\
ConvUtr + Adaptive LGL   & skip1    & 1.34 & 2.43 & 332.45 & 67.02\\
ConvUtr + Adaptive LGL   & skip2    & 1.35 & 2.46 & 328.16 & 67.51\\
ConvUtr + Adaptive LGL   & skip3    & 1.39 & 2.50 & 322.72 & 68.17\\
\hline
\end{tabular}
}
\caption{Ablation study on each blocks.\label{tab:ablation}}
\end{table}

\noindent{\textbf{Skip-connection and adaptive LGL bottleneck:}}
We further investigate the impact of skip connections and adaptive LGL bottleneck on MobileUtr. The results are summarized in Table~\ref{tab:ablation2}.  We first set the fourth layer of the MobileUtr encoder to Global Attention, meaning that both the fourth and fifth layers consist of Transformer blocks. Next, after replacing horizontal skip-connections with downsampling skip-connections, we find that the segmentation performance of MobileUtr is improved and the computational cost is reduced. These ablation results further highlight the necessity of global and local semantic alignment to improve segmentation performance. Finally, after we replace adaptive LGL bottleneck to the fourth layer of the encoder, the segmentation performance is further improved, while the number of parameters and FPS are further reduced. This demonstrates that the adaptive LGL bottleneck plays a joint role in extracting final global information.

\renewcommand{\multirowsetup}{\centering}  
\begin{table}[!t]

\resizebox{1\linewidth}{!}
{
\begin{tabular}{c c| c c c | c}
\hline 

\multicolumn{2}{c}{Network}
& \multicolumn{4}{c}{Metrics(\%)}\\
\cline{0-5}
Encoder & Decoder & Params ↓  & GFLOPs ↓ &  FPS ↑& mIoU ↑\\
\hline
ConvUtr + Global Attention  & horizontal skip3 & 1.67 & 3.34 & 391.08 & 66.50\\
ConvUtr + Global Attention  & skip3 & 1.76 & 2.91 & 416.22 & 67.26\\
ConvUtr + Adaptive LGL  & skip3 & 1.39 & 2.50 & 322.72 & 68.17\\
\hline
\end{tabular}
}
\caption{Ablation study on skip-connection and Adaptive LGL.\label{tab:ablation2}}
\end{table}

\renewcommand{\multirowsetup}{\centering}  
\begin{table}[!]

\resizebox{1\linewidth}{!}
{
\begin{tabular}{c| c c c | c}
\hline 

\multirow{2}{*}{Downsampling strategy} & \multicolumn{4}{c}{Metrics(\%)}\\
\cline{2-5}
& Params ↓  & GFLOPs ↓ &  FPS ↑& mIoU ↑\\
\hline
convolution  & 1.40 & 2.52 & 316.85 & 66.29 \\
maxpooling  & 1.39 & 2.50 & 322.72 & 68.17\\
\hline
\end{tabular}
}
\caption{Ablation study on Downsampling Strategy.\label{tab:ablation3}}
\end{table}

\noindent{\textbf{{Downsampling:}}
Finally, we investigate the impact of different downsampling techniques on medical image feature extraction. We replace all downsampling operations with convolutional downsampling (kernel of 2$\times$2, stride of 2$\times$2), and the ablation results are presented in Table~\ref{tab:ablation3}. We find that segmentation performance is degraded when replaced by convolutional downsampling. This further demonstrates that maxpooling plays an important role in feature extraction from scarce and noisy medical images. It is worth mentioning that we still believe that convolutional downsampling is an important measure for addressing translation invariance in CNNs, However, medical images often exhibit low resolution and minor local edge variations. In comparison to using convolution for downsampling, the conventional pooling operation efficiently filters out the noise present in medical images while maintaining a minimum computational overhead.

\section{Conclusion}
\label{sec:Conclusion}
This paper introduces an innovative medical universal Vision Transformer (ViT) network called MobileUtr. MobileUtr is a groundbreaking ultralightweight network that combines the strengths of CNNs and ViTs. It excels in maintaining low computational complexity, a low parameter count, and high real-time frame rates, while preserving or even enhancing accuracy in general medical segmentation tasks.

The key contribution of MobileUtr lies in its novel fusion concept. This approach enables us to address the challenge of achieving lightweight ViTs while maintaining performance. In comparison to the current state-of-the-art universal medical segmentation network, TransUnet, MobileUtr successfully reduces computational complexity and parameter count by a factor of 10$\times$. Moreover, MobileUtr demonstrates comparable generalization capabilities to state-of-the-art algorithms tailored to specific tasks.

Overall, MobileUtr represents a significant breakthrough as the first successful lightweight implementation of ViTs networks. It achieves state-of-the-art-level accuracy, making it a highly promising and impactful solution for medical image segmentation tasks.

{
    \small
    \bibliographystyle{ieeenat_fullname}
    \bibliography{main}
}
\newpage

\clearpage
\setcounter{page}{1}
\maketitlesupplementary

\section{Discription of Dataset}
The descriptions of each datasets are as follow:

\noindent\textbf{Synapse Dataset}. Synapse multi-organ segmentation dataset~\footnote{https://www.synapse.org/\#!Synapse:syn3193805/wiki/217789}, used for multi-organ CT segmentation, is from the MICCAI 2015 Multi-Atlas Abdomen Labeling Challenge. It comprises abdominal CT scans of 8 organs from 30 cases (3779 axial images). Each CT volume consists of $85 \sim 198$ slices of 512$\times$512 pixels, with a voxel spatial resolution of ($[0.54 \sim 0.54]\times[0.98 \sim 0.98]\times[2.5 \sim 5.0]$) mm$^3$.

\noindent\textbf{BUS Dataset}. The Breast UltraSound (BUS) dataset~\footnote{http://cvprip.cs.usu.edu/busbench/} contains 562 breast ultrasound images collected using five different ultrasound devices, including 306 benign cases and 256 malignant cases, each with corresponding ground truth.

\noindent\textbf{BUSI Dataset}. The Breast UltraSound Images (BUSI) dataset\footnote{https://scholar.cu.edu.eg/?q=afahmy/pages/dataset} collected from 600 female patients, includes 780 breast ultrasound images, covering 133 normal cases, 487 benign cases, and 210 malignant cases, each with corresponding ground truth. Following recent studies \cite{unext, cmunet}, we only utilize benign and malignant cases from this dataset.

\noindent\textbf{TNSCUI Dataset}. The Thyroid Nodule Segmentation and Classification in Ultrasound Images 2020 (TNSCUI) dataset\footnote{https://tn-scui2020.grand-challenge.org/Dataset/} is collected by the Chinese Artificial Intelligence Alliance for Thyroid and Breast Ultrasound (CAAU). It includes 3644 cases of different ages and genders, each with corresponding ground truth.
    
\noindent\textbf{ISIC 2018 Dataset}. The International Skin Imaging Collaboration (ISIC 2018) dataset\footnote{https://challenge.isic-archive.com/data/\#2018} contains 
2,594 dermoscopic lesion segmentation images, each with corresponding ground truth.

\section{Implement Details}
In $LocalAgg$, we use convolution with a kernel size of 9 to achieve local information aggregation. Additionally, in $LocalPro$, we utilize transposed convolution with a kernel of 2 to propagate global context information.

The loss $\mathcal{L}$ between the predicted ${\hat{y}}$ and ground truth $y$ is defined as a combination of binary cross entropy (BCE) and dice loss (Dice):
\begin{equation}
\mathcal{L}=0.5\times BCE\left(\hat{y},y\right)+Dice\left(\hat{y},y\right)
\label{eq}\end{equation}

\section{Segmentation performance with different variants of other method}

The performance comparison between our proposed method MobileUtr and other lightweight model variants are shown in the Table~\ref{tab:supply_other} and Table~\ref{tab:supply_ct}. It can be seen that our proposed method, MobileUtr, achieves best performance. In addition, EMO~\cite{emo} demonstrates relatively excellent results among other method, highlighting the crucial role of the inverted bottleneck design in enhancing representation in segmentation tasks. The inverted bottleneck is also a focal point in our ConvUtr block design. Furthermore, a naive idea is to use EMO bottleneck (iRMB) to replace LGL bottleneck based on the performance in Table~\ref{tab:ablationTransition}. However, as shown in Table~\ref{tab:ablationTransition}, replacing LGL by EMO, results in a significant drop in performance. This indicates that LGL can play a transition role between local (CNN) and global (Transformer), which EMO cannot achieve.  Furthermore, when introducing our adaptive LGL, the performance is further improved, suggesting that adaptive LGL can further enhance the exchange of global and local information flow.

\renewcommand{\multirowsetup}{\centering}  
\begin{table}[!t]

\resizebox{1\linewidth}{!}
{
\begin{tabular}{c c| r r | c}
\hline 

\multicolumn{2}{c|}{Network}
& \multicolumn{3}{c}{Metrics(\%)}\\
\cline{0-4}
Encoder & Decoder & Params↓  & GFLOPs↓ & mIoU↑\\
\hline
ConvUtr + iRMB  & skip3 & 1.23 & 2.39 & 64.18\\
ConvUtr + LGL  & skip3 & 1.36 & 2.49 & 67.32\\
ConvUtr + Adaptive LGL & w/o skip & 1.39 & 2.50 & 68.17\\
\hline
\end{tabular}
}
\caption{Ablation study on Transition Strategy.\label{tab:ablationTransition}}
\end{table}

\renewcommand{\multirowsetup}{\centering}  
\begin{table*}[!]
\resizebox{1\linewidth}{!}
{
\begin{tabular}{r |  r | r | c c | c c | c c | c c}
\hline 

\multirow{3}{*}{Network} & \multirow{3}{*}{Params↓} & \multirow{3}{*}{GFLOPs↓} & \multicolumn{6}{c|}{Ultrasound} &  \multicolumn{2}{c}{Dermoscopy} \\
\cline{4-11}
&&& \multicolumn{2}{c|}{BUS  (\%)} & \multicolumn{2}{c|}{BUSI  (\%)} & \multicolumn{2}{c|}{TNSCUI  (\%)} & \multicolumn{2}{c}{ISIC  (\%)} \\
\cline{4-11}
&&& IOU &F1 & IOU &F1 & IOU &F1 & IOU &F1\\
\hline
\textit{{\color{Gray} light-weight natural network}} &&&&&&&&& \\

MobileViT-xxs~\cite{mobilevit} & 4.30\,M  & 0.55  & 81.62±1.30 & 89.42±1.10 & 63.00±3.04 & 73.71±3.21 & 70.16±0.15 & 80.40±0.16 & 79.76±0.49 & 87.65±0.43\\  

MobileViT-xs~\cite{mobilevit} & 5.77\,M & 1.16 & 82.13±1.20 & 89.69±0.89 & 63.63±3.54 & 74.19±3.51 & 71.13±0.24 & 81.19±0.20 & 79.91±0.49 & 87.73±0.45\\

MobileViT-s~\cite{mobilevit} & 16.49\,M  & 2.71  & 82.57±1.38 & 89.99±1.17 & 64.28±3.78 & 74.68±3.81 & 71.64±0.28 & 81.60±0.25 & 80.12±0.42 & 87.89±0.43\\

EdgeViT-xxs~\cite{edgevits} & 6.83\,M & 0.90 & 81.30±1.65 & 89.12±1.25 & 61.36±3.25 & 72.12±3.26 & 68.36±0.51 & 78.82±0.5 & 78.92±0.63 & 86.97±0.62\\  

EdgeViT-xs~\cite{edgevits} & 10.15\,M & 1.70 & 81.86±1.32 & 89.55±0.95 & 61.94±3.30 & 72.58±3.31 & 69.18±0.26 & 79.44±0.23 & 79.17±0.46 & 87.15±0.46\\

EdgeViT-s~\cite{edgevits} & 10.66\,M & 2.13  & 81.32±1.23 & 89.13±0.99 & 61.12±3.69 & 71.79±4.00 & 68.74±0.29 & 79.19±0.36 & 79.06±0.56 & 87.06±0.55\\

RepViT-m1~\cite{repvit} & 8.48\,M & 1.34 & 76.73±1.12 & 85.98±0.94 & 55.52±2.53 & 67.18±2.73 & 66.61±0.52 & 77.48±0.51 & 78.15±0.60 & 86.43±0.46\\
RepViT-m2~\cite{repvit} & 12.49\,M & 2.09 & 75.93±1.26 & 85.37±1.07 & 54.70±2.21 & 66.27±2.26 & 64.88±0.87 & 75.99±0.75 & 78.03±0.45 & 86.34±0.47\\
RepViT-m3~\cite{repvit}  & 14.37\,M  & 2.79   & 76.51±1.38 & 85.79±1.12 & 56.07±3.51 & 67.62±3.98 & 66.21±0.66 & 77.09±0.49 & 78.34±0.61 & 86.62±0.48\\

EMO-1m~\cite{emo} & 3.36\,M & 0.54 & 84.68±0.96 & 91.44±0.72 & 67.06±3.14 & 77.11±3.15 & 73.01±0.46 & 82.76±0.30 & 80.97±0.48 & 88.46±0.35\\
EMO-2m~\cite{emo} & 4.58\,M & 0.87 & 85.11±1.10 & 91.73±0.75 & 67.90±3.02 & 77.87±2.96 & 73.91±0.39 & 83.39±0.26 & 81.33±0.35 & 88.77±0.28\\
EMO-5m~\cite{emo}  & 7.87\,M & 1.70 & 85.06±0.86 & 91.66±0.63 & 68.27±3.32 & 78.29±3.17 & 74.34±0.56 & 83.77±0.45 & 81.50±0.60 & 88.90±0.48\\
EMO-6m~\cite{emo} & 10.66\,M & 2.13  & 84.89±0.86 & 91.58±0.63 & 67.50±4.26 &77.71±4.23 & 74.02±0.39 & 83.53±0.25 & 81.31±0.73 & 88.74±0.57\\  

\hline
\textbf{MobileUtr}  & 1.39\,M   & 2.51  & \underline{87.28±0.83} & \underline{92.90±0.63} & \underline{72.88±2.72} & \underline{81.18±3.05} & \underline{77.70±0.50} & \underline{85.90±0.41} & \underline{83.23±0.39} & \underline{89.86±0.28}\\
\textbf{MobileUtr-L} & 7.88\,M  & 3.70 & \textbf{87.63±0.91} & \textbf{93.13±0.61} & \textbf{73.91±2.65} &  \textbf{82.16±2.64} & \textbf{78.24±0.38} & \textbf{86.37±0.28}  & \textbf{83.31±0.36} & \textbf{89.88±0.25}\\
\hline
\end{tabular}
}
\caption{Different variants of other method result on Ultrasound and Dermoscopy datasets. \textbf{val} (bold) / \underline{val} (underline) : top method / second method.}
\label{tab:supply_other}
\end{table*}

\section{Visualization results}
We present additional visualization results on five datasets in Fig.~\ref{fig:supply_ct} and Fig.~\ref{fig:supply_other}. As shown in Fig.~\ref{fig:supply_other}, MobileUtr provides more accurate spatial localization and lesion shape for small lesion images (rows 5 and 6). Even in challenging examples with low contrast and unclear boundaries (row 1 and 2), MobileUtr achieves more complete and convex segmentation results. Additionally, for examples under natural light conditions (row 3 and 4), our proposed method demonstrates more precise edges and shapes.

Furthermore, in the task of computed tomography (CT) data segmentation (in Fig.~\ref{fig:supply_ct}), MobileUtr exhibits a strong ability to distinguish the semantics and demonstrates more precise organ localization and segmentation.

\renewcommand{\multirowsetup}{\centering}  
\begin{table*}[htbp]

\resizebox{1\linewidth}{!}
{
\begin{tabular}{r | c | c | c | c c c c c c c c   }
\hline 

\multirow{2}{*}{Network} & \multirow{2}{1cm}{mIoU↑} 
& \multirow{2}{1cm}{Dice↑} & \multirow{2}{1cm}{HD95↓}
& \multicolumn{8}{c}{Synapse (\%)} \\
\cline{5-12}
&&&& Aorta & Gallbladder & Kidney (L) & Kidney (R) & Liver & Pancreas & Spleen & Stomach \\
\hline
\textit{{\color{Gray} light-weight natural network}} &&&&&&& \\
MobileViT-xxs~\cite{mobilevit} & 38.89 & 52.58 & 31.35 & 23.38 & 9.52 & 44.89 & 43.91 & 77.01 & 15.89 & 57.08 & 39.48\\
MobileViT-xs~\cite{mobilevit} & 40.98 & 54.73 & 31.02 & 23.43 & 16.69 & 49.88 & 47.48 & 78.65 & 15.88 & 56.74 & 39.10\\
MobileViT-s~\cite{mobilevit} & 41.66 & 55.57 & 30.81 & 24.63 & 16.58 & 50.94 & 45.85 & 77.74 & 17.90 & 58.80 & 40.83\\

EdgeViT-xxs~\cite{edgevits}   & 35.93 & 49.59 & 36.97 & 20.93 & 10.98 & 38.20 & 36.97 & 75.67 & 14.38 & 51.78 & 38.53\\
EdgeViT-xs~\cite{edgevits}   & 38.13 & 51.80 & 28.71 & 20.98 & 12.37 & 45.93 & 41.25 & 77.07 & 13.93 & 53.46 & 40.04\\
EdgeViT-s~\cite{edgevits}   & 36.86 & 50.68 & 31.82 & 21.97 & 12.23  & 41.42 & 38.96 & 76.43 & 16.00 & 50.71 & 37.16\\

RepViT-m1~\cite{repvit}    & 35.34 & 49.00 & 32.71 & 22.23 & 6.69 & 38.27 & 37.35 & 75.45 & 13.80 & 51.38 & 37.56\\
RepViT-m2~\cite{repvit}    & 34.47 & 47.91 & 32.26 & 19.43 & 7.78 & 36.69 & 37.92 & 74.69 & 13.01 & 49.80 & 36.42\\
RepViT-m3~\cite{repvit}    & 34.07 & 47.61 & 30.69 & 20.71 & 7.16 & 33.56 & 37.28 & 74.33 & 13.28 & 48.13 & 38.09\\

EMO-1m~\cite{emo}   & 45.18 & 59.35 & 32.15 & 28.44 & 20.67 & 51.06 & 48.57 & 80.23 & 22.30 & 59.56 & 50.61\\
EMO-2m~\cite{emo}   & 46.53 & 60.60 & 32.47 & 29.15 & 24.03 & 50.38 & 48.89 & 80.79 & 23.25 & 62.80 & 52.95\\
EMO-5m~\cite{emo}   & 45.10 & 59.08 & 25.00 & 27.00 & 21.87 & 52.57 & 47.89 & 79.76 & 21.93 & 58.76 & 51.02\\
EMO-6m~\cite{emo}   & 45.30 & 59.47 & 27.68 & 28.42 & 22.85 & 51.88 & 48.19 & 79.50 & 22.19 & 60.80 & 48.54\\
\hline
\textbf{MobileUtr}    & \underline{68.17} & \underline{79.13} & \underline{30.96}  &  \underline{79.64} &  \underline{45.96} & \underline{74.93} & \underline{68.69} &  \underline{90.40} & \underline{43.51} &  \underline{81.21} & \underline{60.98}\\
\textbf{MobileUtr-L}  & \textbf{69.09} & \textbf{79.90} & \textbf{26.49} & \textbf{78.99} & \textbf{49.14} & \textbf{72.55} & \textbf{68.29} & \textbf{88.87} & \textbf{50.10} & \textbf{79.57} & \textbf{65.19}\\
\hline
\end{tabular}
}
\caption{Different variants of other method result on CT datasets. \textbf{val} (bold) / \underline{val} (underline) : top method / second method.}
\label{tab:supply_ct}
\end{table*}

\begin{figure*}[t]
  \centering
   \includegraphics[width=0.95\linewidth]{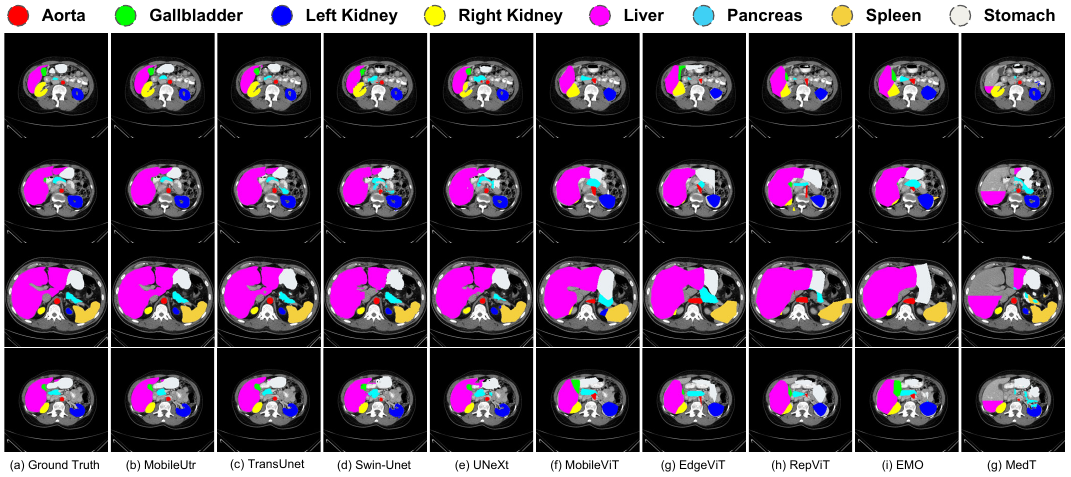}
   \caption{Visualization Results on CT Dataset.}
   \label{fig:supply_ct}
\end{figure*}

\begin{figure*}[t]
  \centering
   \includegraphics[width=0.99\linewidth]
   {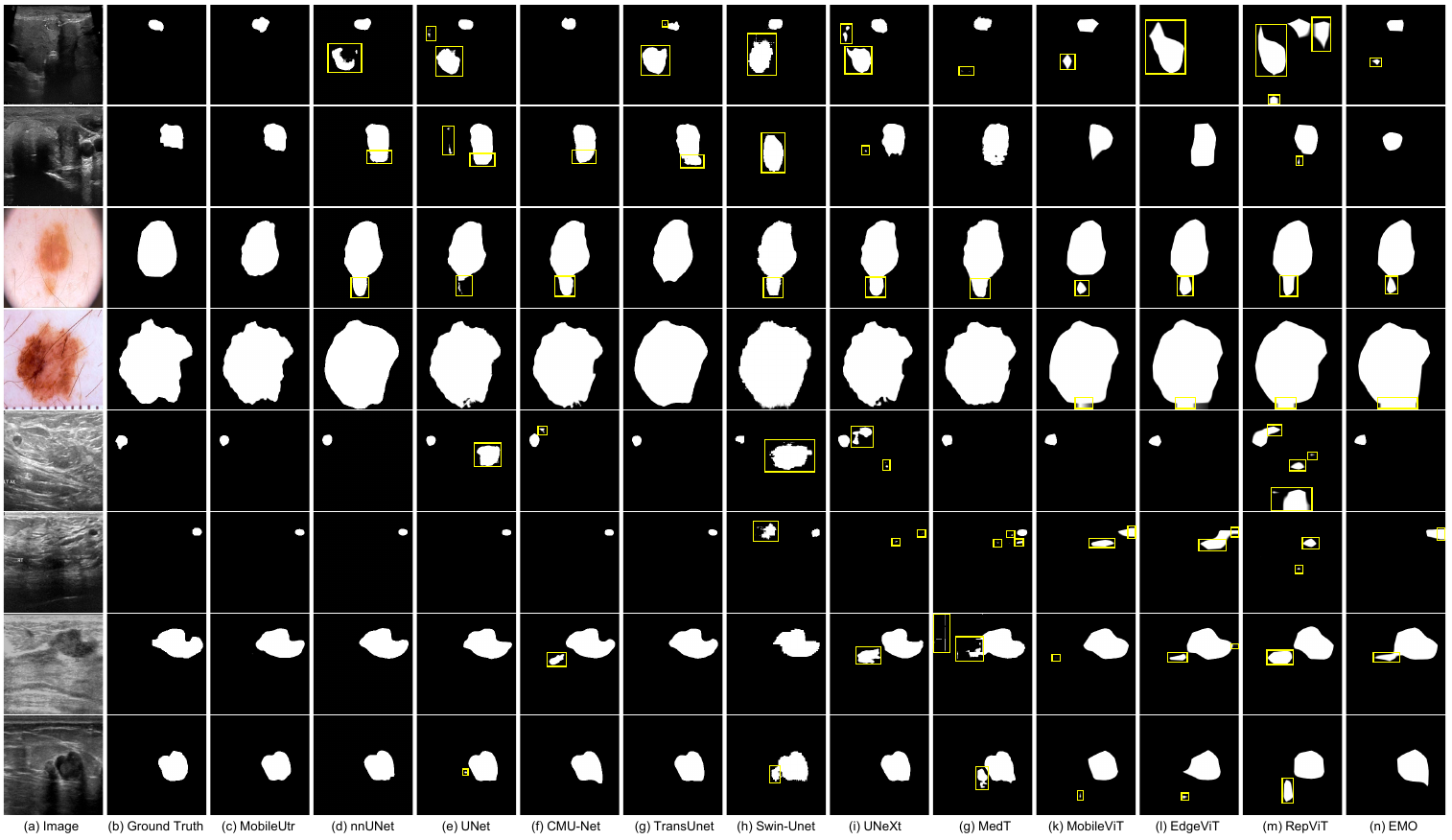}
   \caption{Visualization Results on Ultrasound and Dermoscopy Dataset. Row 1 and 2 - TNSCUI samples, Row 3 and 4 - ISIC18 samples, Row 5 and 6 – BUSI samples, Row 7 and 8 - BUS samples. Yellow boxes represent error segmentation.}
      \label{fig:supply_other}
\end{figure*}

\end{document}